# Exoplanet Biosignatures: Understanding Oxygen as a Biosignature in the Context of Its Environment

Victoria S. Meadows[1,2], Christopher T. Reinhard[3,4], Giada N. Arney[2,5], Mary N. Parenteau[2,6], Edward W. Schwieterman[2,7,8,9,10], Shawn D. Domagal-Goldman[2,11], Andrew P. Lincowski[1,2], Karl R. Stapelfeldt[12], Heike Rauer[13], Shiladitya DasSarma[14], Siddharth Hegde[15,16], Norio Narita[17,18,19], Russell Deitrick[1,2], Timothy W. Lyons[3,4], Nicholas Siegler[20], Jacob Lustig-Yaeger[1,2].

[1] University of Washington Astronomy Department, Seattle, Washington, USA
[2] NASA Astrobiology Institute, Virtual Planetary Laboratory Team, Seattle, Washington, USA
[3] School of Earth and Atmospheric Sciences, Georgia Institute of Technology, Atlanta, Georgia, USA
[4] NASA Astrobiology Institute, Alternative Earths Team, Riverside, California, USA
[5] Planetary Systems Laboratory, NASA Goddard Space Flight Center, 8800 Greenbelt Road, Greenbelt, MD 20771, USA
[6] NASA Ames Research Center, Exobiology Branch, Mountain View, CA 94035, USA
[7] Department of Earth Sciences, University of California, Riverside, California, USA;
[8] NASA Postdoctoral Program, Universities Space Research Association, Columbia, Maryland, USA
[9] NASA Astrobiology Institute, Alternative Earths Team, Riverside, California, USA
[10] Blue Marble Space Institute of Science, Seattle, Washington, USA
[11] Planetary Environments Laboratory, NASA Goddard Space Flight Center, 8800 Greenbelt Road, Greenbelt, MD 20771, USA
[12] NASA Exoplanet Exploration Program, Jet Propulsion Laboratory, 4800 Oak Grove Drive Pasadena, CA 91109
[13] German Aerospace Center, Institute of Planetary Research, Extrasolar Planets and Atmospheres, Rutherfordstraße 2, 12489 Berlin
[14] Department of Microbiology and Immunology, University of Maryland School of Medicine, Institute of Marine and Environmental Technology, University System of Maryland, MD 21202, USA
[15] Carl Sagan Institute, Cornell University, Ithaca, NY 14853, USA
[16] Cornell Center for Astrophysics and Planetary Science, Cornell University, Ithaca, NY 14853, USA
[17] Astrobiology Center, National Institutes of Natural Sciences, 2-21-1 Osawa, Mitaka, Tokyo 181-8588, Japan.
[18] National Astronomical Observatory of Japan, 2-21-1 Osawa, Mitaka, Tokyo 181-8588, Japan.
[19] SOKENDAI (The Graduate University of Advanced Studies), 2-21-1 Osawa, Mitaka, Tokyo 181-8588, Japan
[20] Jet Propulsion Laboratory, California Institute of Technology, 4800 Oak Grove Drive, Pasadena, CA, USA, 91109-8099

***Abstract***: Here we review how environmental context can be used to interpret whether $O_2$ is a biosignature in extrasolar planetary observations. This paper builds on the overview of current biosignature research discussed in Schwieterman et al. (2017), and provides an in-depth, interdisciplinary example of biosignature identification and observation that serves as a basis for the development of the general framework for biosignature assessment described in Catling et al., (2017). $O_2$ is a potentially strong biosignature that was originally thought to be an unambiguous indicator for life at high-abundance. In exploring $O_2$ as a biosignature, we describe the coevolution of life with the early Earth's environment, and how the interplay of sources and sinks in the planetary environment may have resulted in suppression of $O_2$ release into the atmosphere for several billion years, a false negative for biologically generated $O_2$. False positives may also be possible, with recent research showing potential mechanisms in exoplanet environments that may generate relatively high abundances of atmospheric $O_2$ without a biosphere being present. These studies suggest that planetary characteristics that may enhance false negatives should be considered when selecting targets for biosignature searches. Similarly our ability to interpret $O_2$ observed in an exoplanetary atmosphere is also crucially dependent on environmental context to rule out false positive mechanisms. We describe future photometric, spectroscopic and time-dependent observations of $O_2$ and the planetary environment that could increase our confidence that any observed $O_2$ is a biosignature, and help discriminate it from potential false positives. The rich, interdisciplinary study of $O_2$ illustrates how a synthesis of our understanding of life's evolution and the early Earth, scientific computer modeling of star-planet interactions and predictive observations can enhance our understanding of biosignatures and  guide and inform the development of next-generation planet detection and characterization missions.  By observing and understanding $O_2$ in its planetary context we can increase our confidence in the remote detection of life, and provide a model for biosignature development for other proposed biosignatures.

## 1.0 Introduction

The search for life across interstellar distances is a scientific challenge that pushes the boundaries of observational astronomy and requires careful consideration of the signs of life that we will best be able to detect.  These signs are known as 'biosignatures'—life's global impacts on the atmosphere and/or surface of a planetary environment. The most useful biosignatures meet three criteria: reliability, survivability, and detectability, which together enhance the probability that the biosignature can be detected and interpreted as being due to life (e.g., Domagal-Goldman et al. 2011; Seager et al. 2012; Meadows 2017). The reliability criterion seeks to answer whether an observed feature of the planetary environment has a biological origin, and if it is more likely to be produced by life than by planetary processes such as geology and photochemistry.  The survivability criterion determines whether the candidate biosignature gas can avoid the normal sinks in a planetary environment and build up to detectable levels. These sinks include, but are not limited to, destruction by photolysis and photochemistry, reactions with volcanic gases and the surface, and (for soluble gases) dissolution in the ocean.  Finally, the detectability criterion asks whether the gas is spectrally active and clear of overlap with other chemical species in the wavelength region to be observed, and whether it is accessible in high enough abundance and at appropriate regions in the planetary atmosphere to be observed by techniques such as transmission, secondary eclipse, phase curves and direct imaging.

With these three criteria, Earth's abundant molecular oxygen ($O_2$) has been identified as the strongest biosignature for terrestrial planets, and was also initially thought to be straightforward to interpret as being of biological origin. $O_2$ fulfills the three requirements of a biosignature in the following ways:

**Reliability:**  It is the volatile byproduct of the metabolism driven by the dominant source of energy on our planet's surface, oxygenic photosynthesis, and so clearly has a biological origin. Until recently, $O_2$ was considered to have little or no abiotic planetary sources, as the photochemistry driven by the UV spectrum of the Sun produced only trace amounts (e.g. (Domagal-Goldman et al., 2014; Harman et al., 2015), and it did not have significant geological sources, unlike $CH_4$ and $CO_2$, which can be produced by geological processes.  Very early work considered $O_2$ to be an even stronger biosignature when seen along with gases such as $CH_4$ and $N_2O$ in chemical thermodynamic disequilibrium (e.g. Hitchcock and Lovelock, 1967; Lederberg, 1965; Lovelock, 1965; Lovelock and Kaplan, 1975), although given Earth's negligible abiotic $O_2$ sources its high abundance alone indicates a biological origin**.**

**Survivability**:  Indeed, with a strong, planet-wide photosynthetic source, and relatively weak sinks compared to early periods of the Earth's history and different stellar hosts, $O_2$ has risen over time to become the second most abundant gas in our atmosphere (at around 21% by volume).  The atmosphere's more dominant gas, N$_2$—which may also be biologically mediated (Catling and Kasting, 2007; Johnson and Goldblatt, 2015)—is considered a much poorer biosignature because N$_2$ only exhibits strong absorption in the extreme ultraviolet at $\lambda < 0.1$ μm, where several other molecules also absorb and where stars are much less luminous, reducing the reflection or transmission signal from the planetary atmosphere in the UV. Sinks for O$_2$ include outgassing of reduced gases such as CH$_4$ and H$_2$, crustal oxidation, and photolytic destruction in the stratosphere.

**Detectability:**  These factors lead to a high abundance and even mixing throughout the atmospheric column, making O$_2$ accessible to observation by transit spectroscopy.  Depending on the atmospheric composition of a terrestrial, Earth-like planet, transmission spectroscopy may not reach the near-surface environment, instead probing the stratosphere and upper troposphere (Meadows et al., 2017).  This gives O$_2$ a significant advantage over other proposed, often more complex biosignature molecules, including methanethiol and dimethyl disulfide (Pilcher, 2003; Domagal-Goldman et al., 2011; Seager and Bains, 2015), which are more susceptible to photolysis by UV radiation and so are confined largely to the lower troposphere, with significantly smaller concentrations in the stratosphere (Domagal-Goldman et al., 2011).  O$_2$ is also one of the very few biogenic gases which absorbs strongly in the visible and near-infrared, likely wavelength regions to be covered by upcoming NASA missions and ground-based telescopes seeking to characterize extrasolar planets. O$_2$ has strong features at wavelengths < 0.2μm, and a γ band at 0.628 μm, B-band at 0.688 μm, A-band at 0.762μm, and the $a$ $^1\Delta_g$ band at 1.269μm; the A-band is the strongest of these features (Rothman et al., 2013). Additionally, the presence of O$_2$ could be inferred from detection of its photochemical byproduct, O$_3$ (Ratner and Walker, 1972), which has strong bands in the UV (0.2-0.3μm), visible (0.5-0.7μm) and mid-infrared (9.6μm) (Rothman et al., 2013). O$_2$ at high abundance could be identified by collisionally induced absorption as O$_2$ molecules collide with each other or form the O$_4$ complex, which has strong bands in the visible (0.34-0.7um; Hermans et al., 1999; Thalman and Volkamer, 2013) and in the near-infrared (at 1.06μm; Greenblatt et al., 1990, and at 1.27μm; Maté et al., 2000).

However, although O$_2$ is clearly a strong biosignature, recent research has indicated that its availability in a planetary atmosphere and its unique interpretation as a biological product may not be as straightforward as originally thought. These insights have come from exploration of the processes governing the long-term evolution of atmospheric O$_2$ on the early Earth and advances in theoretical modeling of star-planet interactions for planets orbiting other stars.  This research indicates that the picture is far more complex, with both significant false negatives—planetary processes that suppress the buildup of atmospheric oxygen, despite the presence of biological production (Lyons et al., 2014; Planavsky et al., 2014b; Reinhard et al., 2017), and false positives—planetary processes that generate large abundances of oxygen abiotically (Domagal-Goldman et al., 2014; Gao et al., 2015; Harman et al., 2015; Luger and Barnes, 2015; Tian et al., 2014; Wordsworth and Pierrehumbert, 2014), being possible. On the early Earth, it is likely that oxygenic photosynthesis was developed well in advance of the rise of O$_2$ to significant levels in our atmosphere, constituting a false negative. Isotope measurements point to transient low levels of O$_2$ in the early Earth's environment 3.0-2.65 Gya, which were likely generated by oxygenic photosynthesis (Crowe et al., 2013; Czaja et al., 2012; Planavsky et al., 2014a; Riding et al., 2014).  Yet the irreversible, global accumulation of O$_2$ in the atmosphere, which was likely mediated by burial and removal of organic carbon from the Earth's surface environment (Kasting, 2001; Lyons et al., 2014)—evidently occurred somewhat later, between 2.45 - 2.2 Gya (Bekker et al., 2004; Canfield, 2005; Farquhar, 2000).  Recent sulfur isotope data (Luo et al., 2016) place this transition at 2.33 Ga.  However recent studies of Earth's carbon isotope record suggest that the record does not constrain the history of organic burial well enough to evaluate whether changes in fractional organic burial over the last 3.6 Gy can explain the rise of oxygen (Krissansen-Totton et al., 2015).  Other proposed mechanisms for oxygen's rise include hydrogen escape from the upper atmosphere (Catling et al., 2001) and the long-term evolution of volcanic-tectonic processes at Earth's surface (Kump and Barley, 2007; Kasting, 2013; Lee et al., 2016). Additional data suggest that the permanent rise of oxygen to high levels in the Earth's atmosphere (> 1% of the Present Atmospheric Level (PAL)) may have been delayed even further, to 0.8 Gya (Planavsky et al., 2014b), long after the advent of oxygenic photosynthesis.  This delay between the evolution of the biological source of oxygen, photosynthesis, and the subsequent rise of abundant $O_2$ in our atmosphere, is due to diverse geological sinks for $O_2$ in the environment effectively countering its biological production—resulting in a false negative.  This example points to the importance of understanding the characteristics of a planetary environment that may potentially reduce our ability to see a biosignature, even if a strong biological source exists.

Environmental context will also be extremely important for identifying the likelihood for false positives in a planetary environment – abiotic processes that could mimic a biosignature.  $O_2$ was long considered the most robust biosignature possible, because in the Earth's modern environment there were no known false positives, such as geological processes and photochemistry, that could generate it. For example, on Earth, the abiotic production of $O_2$, principally by photolysis of water vapor, is at least a million times less than that produced by photosynthesis (Walker, 1977; Harman et al., 2015).  However, recent modeling studies examining star-planet interactions, especially for exoplanets orbiting M dwarfs, have identified several mechanisms that could potentially lead to the abiotic production of significant amounts of $O_2$.  These mechanisms ultimately rely on conditions that lead to the photolysis of $CO_2$ or $H_2O$, two O-bearing gases that are likely to be common on terrestrial planets in the habitable zone.   One mechanism for $O_2$ buildup relies on the vaporization of oceans, and the subsequent photolysis of water vapor and loss of H, for planets orbiting early, super-luminous M dwarf stars (Luger & Barnes, 2015). A low non-condensable gas inventory in a planetary atmosphere can drive high stratospheric $H_2O$ abundances by lifting the tropospheric "cold trap", leading to the photolysis of $H_2O$ and the loss of H to space; this is another potential abiotic $O_2$ generation mechanism that can affect planets orbiting any type of star (Wordsworth and Pierrehumbert, 2014). Finally, photolysis of $CO_2$ as a source of atmospheric $O_2$ may occur on planets orbiting M dwarfs, and is dependent on both the UV spectrum of the star and the suppression of catalytic or other processes that would allow the CO and O to rapidly recombine (e.g. Tian et al., 2014; Domagal-Goldman et al., 2014; Harman et al., 2015; Gao et al., 2015).   Each of these mechanisms leaves tell-tale signs in the planetary environment that can point to the abiotic source of the $O_2$ (e.g. Schwieterman et al., 2016, 2015b).

Rather than demonstrating that $O_2$ is not suitable as a biosignature, our new and evolving understanding of false positives for $O_2$ *increases* its robustness (Meadows, 2017).  The study of $O_2$ as a biosignature has taught us that by attempting to understand the environmental context, including the possibility for false positives and negatives, we will ultimately improve our interpretation of exoplanet observations and increase our confidence that they do indeed indicate the presence of life.  Knowledge gained from the delayed rise of oxygen during the history of our own planet helps us identify environmental contexts for false negatives for $O_2$, and interdisciplinary modeling studies have helped us to explore and identify the stellar and planetary characteristics that are more likely to lead to $O_2$ false negatives.  This knowledge can guide target selection for habitable planets, and can be used to design observations to search for the necessary environmental context to increase our confidence in the interpretation.   Moreover, these studies serve as a template for future exoplanet biosignature development, and the lessons learned from our study of $O_2$ can help guide the development of a generalized framework for biosignature development and detection.

In Section 2.0, we will describe in more detail the evolution of $O_2$ throughout Earth's history highlighting the ways in which the environment initially suppressed and then finally permitted its development as a detectable biosignature.  In Section 3.0 we will discuss the known potential false positive mechanisms for terrestrial exoplanets, and their likely observable environmental impact.  In Section 4.0 we will discuss the specific observational implementation for $O_2$ detection and how confidence in its biogenicity can be enhanced, including (1) characterization of planetary, stellar and planetary system environmental context; (2) the search for $O_2$; and (3) the specific environmental characteristics that could be sought to help discriminate false negatives and positives. We will describe the photometric and spectroscopic measurements required to execute these three steps for different near-term exoplanet observing facilities, including JWST, extremely large ground-based telescopes, and for proposed space-based direct imaging coronographic/starshade missions.  In Section 5.0 we will discuss and prioritize the most important discriminating measurements, and then discuss the robustness of any detection as a function of accessible wavelength range and other telescope characteristics.  In Section 6.0 we will present our conclusions.

## 2.0 The Co-Evolution of Life with its Environment, False Negatives, and the Rise of Earth's $O_2$.

Whether $O_2$ rises to detectable levels in a planetary atmosphere will depend on the balance between sources and sinks for $O_2$ in surface environments and in particular whether planetary processes, such as geological burial of organic carbon (e.g., Des Marais et al., 1992), atmospheric escape of hydrogen (Catling et al., 2001), or changes in reductant fluxes to the surface (Kasting, 2013) can support the buildup of biologically-generated $O_2$ in the atmosphere on long timescales.

On Earth, the principal source of $O_2$ is oxygenic photosynthesis, which uses energy from the Sun to convert carbon dioxide and liquid water to glucose, with oxygen as a byproduct (Eq. 1).

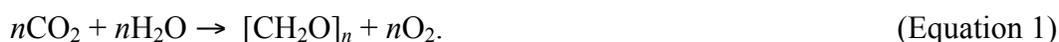

$$n\mathrm{CO}_2 + n\mathrm{H}_2\mathrm{O} \;\rightarrow\; [\mathrm{CH}_2\mathrm{O}]_n + n\mathrm{O}_2. \qquad\qquad\qquad (\text{Equation 1})$$

These globally available and energetically-rich components of the habitable terrestrial environment allow oxygenic photosynthesis to dominate large fractions of the planetary surface environment (Kiang et al., 2007a; Léger et al., 2011), and it is critical in maintaining much of the living biomass on our planet today (Falkowski et al., 2000). Its dominance in our biosphere and its global distribution have also made photosynthesis more detectable on a planetary scale. In this section, we will discuss the evolution of oxygenic photosynthesis, the suppression and eventual rise of atmospheric $O_2$, and the interplay between biological production of $O_2$ and its consumption by sinks present in early Earth surface environments.

## 2.1 The evolution of oxygenic photosynthesis

There are tantalizing suggestions that life may have emerged very early during the ~4.5 billion year evolution of our planet. Carbonaceous compounds enriched in $^{12}C$ (Mojzsis et al., 1996b; Rosing, 1999), potentially biogenic sedimentary structures (Nutman et al., 2016), and purported microfossils (Dodd et al., 2017) are present in the oldest sedimentary rocks on Earth. Furthermore, $^{12}C$-enriched graphitic inclusions in recycled zircons have been suggested to date the emergence of life to before the earliest sedimentary rocks, at ~4.1 billion years ago (Ga; Bell et al., 2015). These observations are consistent with a biological origin, but in all cases are somewhat equivocal (see, for example, van Zuilen et al., 2002; Lepland et al., 2002; McCollom & Seewald, 2006). Nevertheless, strong isotopic evidence for the presence of important microbial metabolisms such as methanogenesis (Ueno et al., 2006) and dissimilatory sulfate reduction (Shen et al., 2001; Ueno et al., 2009) is observable by at least ~3.5 Ga, suggesting that although the precise timings of biochemical steps remains to be elucidated, key landmarks in evolution of life—the emergence of life from prebiotic chemistry, the development cellularity, and the development of metabolism—likely occurred relatively soon after Earth's formation.

The evolution of oxygenic photosynthesis and the resulting oxygenation of the atmosphere and oceans was arguably one of the most important biological innovations on Earth. This rise of atmospheric oxygen ultimately allowed for subsequent biological (Knoll et al., 2006) and mineralogical diversification (Hazen et al., 2008). The light reactions of oxygenic photosynthesis in cyanobacteria, algae, and plants use two linked photosystems: photosystems I (PSI) and II (PSII). This linkage allows for the extraction of an electron from water, which has a very positive redox potential ($O_2/H_2O$ pair; $E_0' = + 0.87$ V) (see overview in Schwieterman et al., this issue). This configuration of photosynthetic machinery is very complex, and the evolution of cyanobacteria performing oxygenic photosynthesis was preceded by more primitive anoxygenic phototrophs (e.g., Xiong et al., 2000).

The photosynthetic machinery of anoxygenic phototrophs is much less complex. These primitive phototrophs use a single photosystem, which can be classified into one of two families—type I reaction centers, including photosystem I (PSI) in chloroplasts, cyanobacteria, green sulfur bacteria, heliobacteria, and phototrophic Acidobacteria, and type II reaction centers, such as photosystem II (PSII) in chloroplasts, cyanobacteria, purple bacteria, and green non-sulfur bacteria (Allen & Williams, 1998; Bryant et al., 2007). Rather than using chlorophyll pigments that absorb light at wavelengths in the visible range, anoxygenic phototrophs utilize a diversity of bacteriochlorophyll pigments that absorb in the near-infrared (Senge & Smith, 1995).

One school of thought on the evolutionary transition from anoxygenic to oxygenic photosynthesis emphasizes the importance of the redox potential of electron donors or reductants. The most ancient form of anoxygenic photosynthesis likely used strong reductants such as H$_2$ or H$_2$S donating electrons to PSI (Olson and Pierson, 1986). These volcanogenic compounds were likely in limited supply on the early Earth, providing evolutionary pressure for the development of a second photosystem (PSII) that could utilize weaker but more abundant reductants, such as Fe(II) and water (Olson, 1970, 1978).

For cyanobacteria, it has been hypothesized that there was a step-wise increase in the redox potentials of reductants donating electrons to PSII (Pierson and Olson, 1989; Olson, 2006). Modern cyanobacteria produce oxygen as a waste product resulting from the oxidation of water, but this may not have always been so. There is a large difference in the redox potentials between the O$_2$/H$_2$O pair (E$_0$' = + 0.87 V) of oxygenic photosynthesis and the S$_0$/HS$^-$ pair (E$_0$' = - 0.27 V) commonly used by anoxygenic photosynthesis. Researchers have speculated that an intermediate reductant, such as Fe(II) (E$_0$' = + 0.30 V at circumneutral pH), could have bridged the gap and acted as a transitional electron donor before water (Pierson and Olson, 1989; Olson, 2006). The widespread abundance of reduced iron on the early Earth would have made it particularly suitable as an electron donor for photosynthesis. Pierson et al. (1993) point out that the oxidized iron products of this type of photosynthesis would have provided substantial protection from UV radiation for surface-dwelling phototrophs prior to the development of an ozone shield.

Due to the high redox potential of water, the energy yield of oxygenic photosynthesis is much greater than that of anoxygenic photosynthesis. This greater energy yield, coupled with the wide availability of water as an electron donor compared to the limited availability of volcanogenic reductants, allowed oxygenic phototrophs to spread easily and dominate aquatic and terrestrial habitats. Indeed, the evolution of oxygenic photosynthesis would ultimately result in a radical restructuring of Earth surface environments, with significant impacts on all major biogeochemical cycles. The so-called Great Oxidation Event, which occurred at ~2.3 Ga (Bekker et al., 2004; Luo et al., 2016), set the stage for the evolution of more complex life forms dependent on oxygen. Consequently, oxygen is not only a promising biosignature, but it may also indicate an environment in which multicellularity, and even intelligence, can be supported.

It is of course difficult to estimate the likelihood that oxygenic photosynthesis would evolve on an exoplanet. However, if volcanism is a common geologic process on rocky habitable planets, and if water and CO$_2$ are widespread and abundant, then the same basic selective pressures that gave rise to oxygenic photosynthesis on the early Earth would be operative. The possibility thus exists that the same evolutionary pressures based on electron donor availability could play out on an exoplanet, leading to the same metabolic outcome.

If photosynthesis did evolve on an exoplanet, it would likely co-evolve with its environment to develop light-harvesting pigments that make use of the peak solar photon flux onto the planetary surface (Kiang et al., 2007a,b), which could potentially be predicted. The peak photon flux will be the result of the interplay between the incident solar spectrum and subsequent absorption by molecules in the atmosphere, as well as efficiency considerations for near-infrared radiation, which may be too low in energy to drive photosynthesis efficiently beyond a limiting wavelength (Kiang et al., 2007a,b). On Earth, ozone absorption in our atmosphere drives the Sun's incident peak photon flux near 600nm redward. The peak photon flux at the surface is at 688nm, where the principal photosynthetic pigment, chlorophyll *a*, absorbs (Kiang et al., 2007b). However, on the early Earth, without abundant ozone, the peak photon flux at the surface may have been different, but potentially also pushed redward by strong blue absorption from hydrocarbon haze. Such hazes may have been present during early periods of Earth's history when the atmosphere was more reducing (Arney et al., 2016). In the case of exoplanets the spectrum of the host star will be well known in the visible range, and the planetary atmospheric composition may be constrained by spectral observations. These observations will allow radiative transfer models to predict photon fluxes at the planetary surface, and potentially help identify the most likely spectral region for photosynthetic pigment absorption.

## 2.2 The suppression, and ultimate rise, of oxygen on the early Earth.

Although the evolution of oxygenic photosynthesis represented a critical evolutionary innovation, it was not a sufficient condition for appreciable and persistent planetary oxygenation, which may not have happened until as recently as 0.8 Ga (Lyons et al., 2014). To see why this may have been so requires a basic discussion of the factors modulating Earth's global $O_2$ cycle (Figure 1). Oxygenic photosynthesis on the modern Earth fixes carbon (i.e., converts atmospheric $CO_2$ into organic carbon) at a net rate of ~100 Pg ($10^{15}$ grams) of carbon per year, distributed roughly evenly between the marine and terrestrial photosynthetic biospheres (Field et al., 1998). This process releases a stoichiometric amount of free $O_2$ to the environment, at a ratio of ~1:1. However, the vast majority of this $O_2$ is consumed very rapidly through aerobic respiration in surface environments, such that only a very small fraction of this (less than 1%) escapes respiration and is ultimately buried in marine and terrestrial sediments (Berner, 1982)

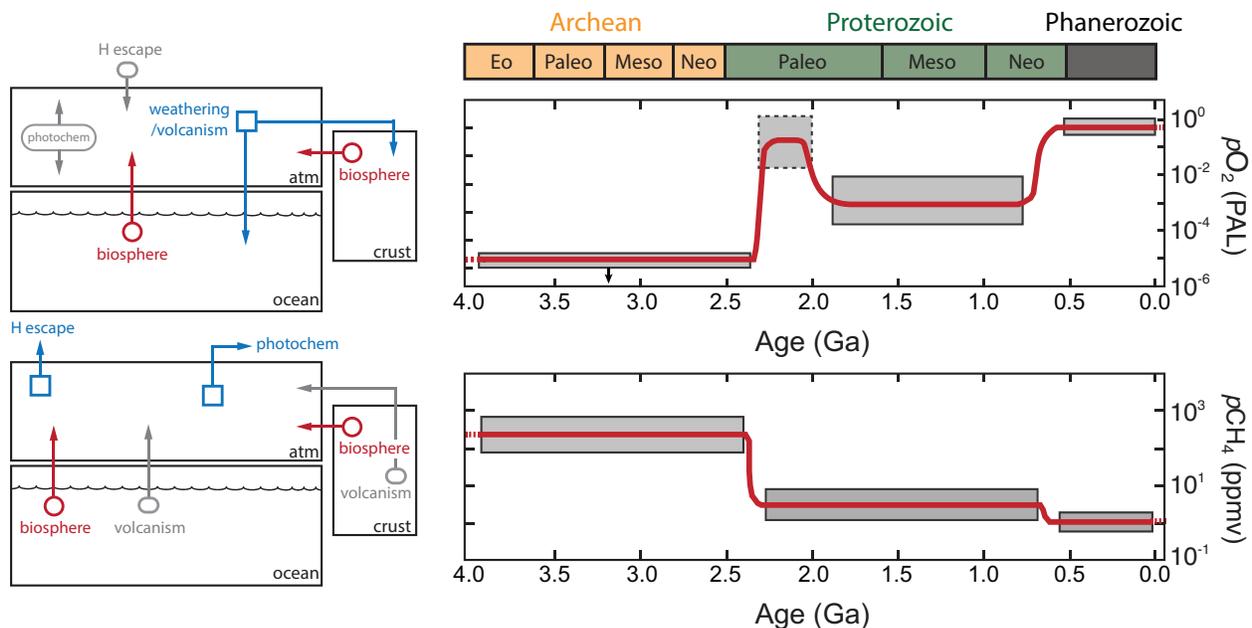

*Figure 1 Overview of the major processes controlling atmospheric $O_2$ and $CH_4$ levels (left panels). Red circles/arrows denote net biogenic sources (e.g., after accounting for biological consumption), grey ovals/arrows denote potential abiotic sources, and blue squares/arrows denote net sink fluxes. Schematic depiction of the evolution of atmospheric $O_2$ and $CH_4$ throughout Earth's history (right panels). Shaded boxes show ranges based on geochemical*

*proxy or model reconstructions, while red curves show possible temporal trajectories through time. (credit: Chris Reinhard)*

The relatively slow 'leak' of organic carbon into the Earth's upper crust, together with the co-occurring burial of reduced sulfur as sedimentary pyrite ($FeS_2$), represent net sources of $O_2$ to Earth's atmosphere. These fluxes are augmented by the escape of $H_2$ to space, which scales with the total abundance of H-bearing reducing gases at the homopause (e.g., $CH_4$ and $H_2$) and is thus a very small flux on the modern Earth (Catling et al., 2001). Input fluxes of $O_2$ are balanced on long timescales by $O_2$ consumption during the uplift and oxidative weathering of reduced mineral phases in Earth's crust (including previously buried organic carbon and pyrite sulfur) and through reaction with volcanic and metamorphic reductants (Kasting and Canfield, 2012).

Whether a planet develops a biogenic $O_2$-rich atmosphere will thus depend not only on the evolution of oxygenic photosynthesis but also on the long-term balance between sources and sinks of $O_2$ at the surface (e.g. Gebauer et al., 2017). The accumulation and stability of large biogenic $O_2$ inventories in a planetary atmosphere will in turn depend on how input/output fluxes scale with atmospheric $O_2$ levels (if at all) and on secular changes in planetary degassing, differentiation, and atmospheric escape. An important consequence of this is that the evolution of oxygenic photosynthesis cannot, on its own, be expected to yield a remotely detectable oxidizing atmosphere. The planetary environment also plays a crucial role.

Earth's history provides an instructive example. Constraints from stable isotopes and trace element proxies indicate the evolution of oxygenic photosynthesis on Earth by at least ~3.0 Ga (Planavsky et al., 2014a) and perhaps much earlier (Rosing & Frei, 2004). However, complementary isotopic constraints indicate that Earth's atmosphere was pervasively reducing until ~2.5 Ga (Farquhar et al., 2001; Pavlov & Kasting, 2002; Zahnle et al., 2006). There thus appears to have been a very significant period on Earth during which oxygenic photosynthesis was present but large amounts of $O_2$ did not accumulate in Earth's atmosphere. In addition, there is some evidence that after the initial accumulation of $O_2$ in Earth's atmosphere at ~2.3 Ga atmospheric $O_2$ levels remained relatively low for much of the subsequent ~2 billion years (Planavsky et al., 2014b; Cole et al., 2016; Tang et al., 2016), during which biogenic $O_2$, though clearly present, may have been challenging to detect remotely given current technology (Reinhard et al., 2017) .

In sum, the emergence of a biogenic $O_2$-rich atmosphere will depend on both the evolution of oxygenic photosynthesis, as well as geochemical dynamics at the planetary surface that are favorable for the long-term accumulation of a large atmospheric $O_2$ inventory: If planetary conditions are not favorable, then a false negative will occur.  These dynamics will, in turn, depend on a series of planetary factors that may be challenging to constrain observationally or from first principles. For example, heat flux from a planetary interior (as constrained by radiogenic element inventory and planet size), oxygen fugacity of the planetary mantle (as constrained by both initial chemistry and long-term recycling of materials from the surface), the degree of crustal differentiation (as constrained by both overall heat fluxes and planetary rheology), and ocean chemistry can interact to buffer atmospheric $O_2$ to low levels despite the presence of oxygenic photosynthesis. The ability to constrain these contextual variables via observations of the planet and star, or via or modeling may ultimately form a critical component of diagnosing false negatives for $O_2$ on living planets.

## 2.3 Earth's history of $O_2$'s companion biosignatures $CH_4$ and $N_2O$

The role of the environment in enhancing or suppressing the atmospheric accumulation of biogenic gases also applies to other species, such as methane ($CH_4$) and nitrous oxide ($N_2O$), and the detection of these gases could be used to strengthen confidence that the observed $O_2$ in an exoplanet atmosphere is indeed biogenic (e.g. Hitchcock & Lovelock, 1967). Atmospheric levels of these gases will be a function of production at the surface (whether biological or abiotic), but will also depend strongly on sink fluxes due to biological recycling, geologic processes, and photochemistry in the atmosphere. For example, $N_2O$ is photochemically unstable in reducing atmospheres (Roberson et al., 2011), making it difficult to sustain high atmospheric levels without large surface fluxes. The atmospheric lifetime and hence abundance of $N_2O$ depends on the spectrum of the host star, with M dwarfs allowing longer lifetimes and higher abundances for the same surface flux (Segura et al., 2005; Rauer et al., 2011; Rugheimer et al., 2013, 2015). However, in general, relatively high $N_2O$ fluxes will be required to maintain elevated atmospheric $N_2O$ levels unless the atmosphere is already oxidizing, with high $O_2/O_3$ or $CO_2$ levels. These kinetic constraints on biogenic gas abundance in planetary atmospheres have led to approaches that emphasize the magnitude of chemical disequilibrium between species (e.g., $O_2/O_3$ and $CH_4$), and this approach can potentially provide more specific information about the magnitude of production/consumption fluxes compared to the abundance of a single biogenic gas (Lovelock and Kaplan, 1975; Kasting et al., 2013; Krissansen-Totten et al., 2015).

The evolution of atmospheric chemistry on Earth illustrates how searching for secondary gases to strengthen $O_2$'s biogenic origin may be both challenging and illuminating (Fig. 3):

**The Archean (~4.0-2.5 billion years ago, Ga) – low $O_2$, high $CH_4$**: During most of the Archean Eon, background atmospheric $O_2$ levels were vanishingly low (likely less than $\sim 10^{-7}$ times PAL; Zahnle et al., 2006), with peak $O_3$ levels correspondingly low (Kasting and Donahue, 1980). Theoretical models predict that under such reducing atmospheric conditions, and with relatively low oxidant levels in the ocean (e.g., dissolved $O_2$, $NO_3^-$, $SO_4^{2-}$), atmospheric $CH_4$ levels may have been ~2-3 orders of magnitude above those of the modern Earth (Fig. 3; Pavlov et al., 2003). Indeed, new isotopic tools are emerging for fingerprinting the photochemical impacts of much higher atmospheric $pCH_4$ and its modulation by Earth's biosphere (e.g., Izon et al., 2017). Although there is now firm evidence for transient increases in atmospheric $O_2$ levels during the late Archean (Anbar et al., 2007; Kendall et al., 2010), the magnitude and duration of these $O_2$ pulses is not well-constrained at present. Thus, for much of Archean time Earth's atmosphere was characterized by very low $O_2/O_3$ but elevated $CH_4$.

**The Paleoproterozoic (~2.2-2.0 Ga) – higher $O_2$, low $CH_4$, possible $N_2O$:** Following the first large-scale oxygenation of the ocean-atmosphere system after ~2.3 Ga, there was a period of the Paleoproterozoic during which atmospheric $O_2/O_3$ levels may have risen to (or even exceeded) modern levels for perhaps 100-200 million years or more (Fig. 3; Lyons et al., 2014). This increase would, in all likelihood, have led to a corresponding drop in atmospheric $CH_4$ levels, which may have photochemically stabilized $N_2O$ for the first time in Earth's history.

**The mid-Proterozoic (~1.8-0.8 Ga) – lower O₂, low CH₄**: Atmospheric $O_2/O_3$ levels appear to have dropped markedly and remained at relatively low background levels for much of the mid-Proterozoic (Lyons et al., 2014; Planavsky et al. 2014, Cole et al., 2016; Reinhard et al. 2016; Tang et al., 2016). At the same time, recent Earth system modeling of the $CH_4$ cycle under mid-Proterozoic conditions as inferred from the geochemical proxy record (e.g., atmospheric $pO_2$ and marine $SO_4^{2-}$ levels) imply that atmospheric $CH_4$ may also have been rather low, on the order of ~1-10 ppmv (Olson et al., 2016).

**The late Proterozoic (~800-550 million years ago (Ma)) – higher O₂, low CH₄, higher N₂O:** The geochemical record indicates another series of significant shifts in ocean-atmosphere redox during the late Proterozoic, after which atmospheric $O_2/O_3$ levels may have ultimately stabilized at roughly modern levels. Atmospheric $pO_2$ subsequently varied between ~0.25-1.5 PAL over the last ~500 million years (Berner et al., 2006). Earth's atmosphere during this period has been characterized by correspondingly low atmospheric $CH_4$ (Fig. 3) and presumably relatively high atmospheric $N_2O$.

One striking implication of this history is that throughout Earth's evolution, either $O_2$ or $CH_4$, or neither, dominated the atmosphere, and so the canonical $O_2$-$CH_4$ disequilibrium biosignature may have been challenging to observe in the atmosphere over most of Earth history. This is despite the fact that oxygenic photosynthesis has been present at Earth's surface for at least ~3 billion years (Planavsky et al.,. 2014a), and methanogenesis has been extant for ~3.5 billion years (Ueno et al., 2006). The source-sink relationships for the coupled $O_2$ and $CH_4$ cycles throughout Earth's history may thus have led to a persistent false negative signal, during which major biogenic gases were being produced and consumed at potentially high rates, but did not co-accumulate in the atmosphere at potentially detectable levels (Reinhard et al., 2017). Possible exceptions include relatively high $O_2/O_3$ during the Paleoproterozoic and during the last ~500 million years, and potentially detectable $O_3$ during the mid-Proterozoic (Reinhard et al., 2017). However, the possibility remains that Earth's surface biosphere, definitively established by at least ~3.5 billion years ago, may not have been detectable until relatively recently in Earth's geologic history.

## 2.4 Secondary Photosynthetic Biosignatures

The Earth also demonstrates alternative ways in which our photosynthetic biosphere has impacted the environment beyond the abundant oxygen that eventually rose in our atmosphere. These secondary biosignatures include surface reflectivity signatures from both light harvesting (Kiang et al., 2007a, 2007b) and other pigments developed for non-photosynthetic purposes by phototrophs (Hegde et al., 2015; Schwieterman et al., 2015a), as well as the strong jump in reflectivity from the red edge of vegetation (Gates et al., 1965). That increase in reflectivity in Earth's spectrum typically occurs just past chlorophyll absorption near 0.7um (Kiang et al., 2007a). Another secondary biosignature of photosynthesis in globally-averaged spectra of our planet is the seasonal variations in the abundance of $CO_2$ (Meadows, 2008) due to growth and decay of vegetation on land which imparts a ~2% seasonal variation in $CO_2$ abundance (Keeling 1960; Hall et al. 1975; Keeling et al. 1976). Earlier in Earth's history before the permanent rise of oxygen, small amounts of biogenic $O_2$ may also have shown seasonal variation (Reinhard et al., 2016; see Schwieterman et al., this issue, for a review).  Either a corroborative surface or temporal signature may be sought as a secondary confirmation of a photosynthetic origin for any detected $O_2$ in a planetary atmosphere.  The vegetation red edge has likely been widespread on Earth since about 0.46 Ga (Carroll, 2001; Igamberdiev and Lea, 2006), corresponding with the rise of land plants ,and may be coincident with the rise of $O_2$ to near-modern levels.  However, in false negative situations where planetary processes are suppressing the rise of photosynthetically-generated $O_2$, surface reflectivity biosignatures from photosynthetic and non-photosynthetic pigments produced by phototrophs may have been the only way to detect photosynthesis (Schwieterman et al., 2015a).

## 3.0  Star-Planet Interactions, and the Generation of False Positives for $O_2$ in Exoplanetary Environments

While false negatives may reduce the atmospheric signal and so preclude the detectability of biosignatures, false positives complicate the interpretation of any $O_2$ that is observed.  False positives for $O_2$ on uninhabitable planets have been postulated for decades (Schindler & Kasting, 2000; Selsis et al., 2002; Des Marais et al., 2002 and references therein), but were more likely to occur for planets outside the habitable zone, and so were considered easy to identify and exclude.  However several plausible $O_2$ false positive scenarios for planets in the habitable zone have since been identified (e.g. Wordsworth & Pierrehumbert, 2014; Luger & Barnes, 2015; Tian et al, 2015: Harman et al., 2015; see Meadows, 2017 for a more detailed review).  These mechanisms rely primarily on conditions that lead to the photolysis of $H_2O$ and/or $CO_2$ in a terrestrial planet atmosphere, and in some cases stabilize the photolytic byproducts of $CO_2$ from recombination by suppressing the action of catalysts.   Rather than weakening oxygen's role as a biosignature, knowledge of false positives allows us to identify observable characteristics of the stellar or planetary environment that indicate the mechanisms that generate them.   Each potential false positive mechanism that is ruled out by observations increases the robustness of a biogenic interpretation for $O_2$.   Below we summarize some of the most significant false positive mechanisms for planets in the habitable zone.

### 3.1 Low Non-Condensable Gas Inventories

The possible abiotic $O_2$ generation mechanism that will likely be the most difficult to observationally recognize and preclude involves photolysis of water -  and subsequent hydrogen loss - from terrestrial atmospheres that are depleted in non-condensable gases such as $N_2$ (Wordsworth & Pierrehumbert, 2014). This mechanism may produce Earth-like quantities of $O_2$ on an ocean-bearing world.  To work, the effectiveness of the "cold trap" – the rapid reduction in temperature with altitude that causes rising water vapor on Earth to condense and thereby remain trapped in the troposphere – must be weakened.   This occurs when the atmospheric temperature is high, or when the total inventory of non-condensable gases – for example, $N_2$, $O_2$ or $CO_2$ – is low.  Because this mechanism relies primarily on a planetary property, a low inventory of gases that do not condense at typical habitable zone terrestrial planetary temperatures and pressures, it may work for planets orbiting stars of any spectral type, including Sun-like stars. In these cases, water rises into the stratosphere where it is more vulnerable to photolysis by incident stellar UV radiation.  The water's H atoms escape, leaving oxygen to build up abiotically in the atmosphere.  This continues until $O_2$, itself a non-condensable gas, overwhelms surface sinks to reach a sufficiently high abundance that it can establish a cold trap, and halt the loss of water vapor.   For an abiotic planet that is largely Earth-like except for the N$_2$ inventory, Wordsworth & Pierrehumbert (2014) calculate that this proposed mechanism could result in scenarios where a planet that exhibited a very Earth-like O$_2$ partial pressure (~ 0.15bars), while losing significant amounts of water.

## 3.2 Enhanced M dwarf Pre-Main Sequence Stellar Luminosity

A mechanism first proposed by Luger & Barnes (2015) focuses on the effects of the pre-main-sequence, super-luminous phase of young M dwarf stars on terrestrial planet environments, and this mechanism may be capable of abiotically generating thousands of bars of atmospheric O$_2$. Before settling in to their main sequence hydrogen burning phase, the extended contraction phase of young stars makes them significantly more luminous (e.g Baraffe et al., 1998) Planets that form in what will become the main-sequence habitable zone are subjected early on to very high levels of radiation (Lissauer, 2007). This super-luminous phase is longer for smaller mass M dwarfs when compared to other stellar spectral types, and can extend for up to 1Gyr (Baraffe et al., 1998). Models suggest that this super-luminous phase can drive the loss of up to several Earth ocean equivalents of water due to evaporation and hydrodynamic escape (Luger and Barnes, 2015; Tian, 2015) for an M dwarf terrestrial planet that forms within what will become the main sequence habitable zone. The resultant water-rich atmosphere will be susceptible to photolysis and hydrogen escape, thereby producing potentially hundreds or thousands of bars of O$_2$ (Luger & Barnes, 2015; photolysis of an Earth ocean of water can produce ~240 bars of O$_2$, Kasting, 1997).

The amount of O$_2$ generated in this scenario is a strong function of stellar spectral type, XUV flux, original water inventory, planetary mass, and the position of the planet in the habitable zone.  Planets in the outer regions of the habitable zone for M0-M3V stars are the least likely to generate significant amounts of abiotic O$_2$ (Luger & Barnes, 2015), along with M1-M3V dwarfs with lower stellar XUV, as slower water loss may preclude O$_2$ generation and buildup (Tian, 2015).   However, for planets orbiting later type M dwarfs (M4V and later) depending on the planet's original water inventory, mass, stellar parameters and the strength of surface sinks (Rosenqvist and Chassefière, 1995; Schaefer et al., 2016), up to several hundreds of bars of photolytically-produced O$_2$ could build up in the atmosphere (Luger & Barnes, 2015).   For the recently discovered Proxima Centauri b, orbiting an M5.5V star, estimates of ocean loss during the 169 +/- 13 Myr (Barnes et al., 2017) before the planet enters the habitable zone are sensitive to the assumed form of the XUV evolution.   Results for global water loss range from less than one ocean (<250 bars O$_2$ generated), assuming a constant XUV flux for the host star for the first 3 Gyr (Ribas et al., 2016), up to 3-10 oceans lost (750-2500 bars of O$_2$ generated), assuming that the XUV scales with the decreasing bolometric luminosity (Barnes et al., 2016). If substantial CO$_2$ is also outgassed over time, and the planet has little or no surface ocean in which to sequester CO$_2$, then these types of worlds could sustain atmospheres with large quantities of both CO$_2$ and O$_2$ (Meadows et al., 2017).  Over time the O$_2$ may be lost and CO$_2$ may continue to accumulate.   Although we have evidence that Venus lost a global "ocean" that was at least 3 m deep (deBergh et al., 1991) any abiotic O$_2$ generated was subsequently lost to surface or interior sinks (Rosenqvist & Chassefieré et al.,1995; Schaefer et al., 2016) or top-of-atmosphere loss processes. These may include the recently discovered "electric wind" surrounding Venus, which can strip oxygen ions from the planet (Collinson et al. 2016).

## 3.3. Stellar Spectrum Driven Photochemical Production

Another mechanism for formation of abiotic $O_2$ and its proxy, $O_3$, is planetary photochemistry, especially of $CO_2$. Each photochemical reaction requires photons that exceed a threshold energy level - or equivalently, photons at wavelengths shorter than a threshold wavelength - to be absorbed by and split the molecule undergoing photolysis.   Planetary photochemistry is therefore strongly sensitive to the UV spectral energy distribution of the parent star  (e.g. Segura et al., 2003, 2005; Grenfell et al., 2007, 2014; Rugheimer et al., 2013; 2015;), and in particular the ratio of shorter to longer wavelength UV radiation, which can split molecules and drive reactions that change the composition of the atmosphere, without relying on atmospheric escape.

The ultimate composition of the planetary atmosphere will then depend on the sources and sinks for photochemical reactions provided by the planetary environment. For example, photochemical abiotic $O_2$ production will depend strongly on the source of O atoms, which is primarily controlled by the abundance and photolysis rates of atmospheric $CO_2$, $SO_2$, $H_2O$ and other O-bearing gases.   Similarly, the atmospheric sink will depend on the availability of H atoms in the atmosphere from $H_2$, $H_2S$ and hydrocarbons.   While water vapor can be sourced from a planetary ocean, the availability of other gases, such as $CO_2$, $SO_2$ and reducing (H-bearing) gases, are governed by the planet's volcanic outgassing rates, which can be sustainable on long geological timescales.   In particular, $CO_2$ is likely a common atmospheric gas on terrestrial planets.     In our own Solar System, $CO_2$ dominates atmospheric composition on Mars and Venus, and was likely a significant component of the early Earth's atmosphere as well (Kasting 1993; Sheldon et al. 2006; Sleep, 2010; Driese et al. 2011).   The key reactions to produce $O_2$ from $CO_2$ photolysis involve $CO_2 + hv \rightarrow CO + O$, and $O + O + M \rightarrow O_2 + M$.  The yield of $O_2$ from this process depends on how efficiently the back reaction to regenerate $CO_2$ can occur. This in turn depends on the atmospheric abundance of catalysts such a $HO_x$ (e.g. from water vapor; Tian et al., 2014) or $NO_x$ (e.g. from cosmic rays; Grenfell et al., 2013) (c.f. Yung and DeMore, 1989; Stock, et al. 2017).

Aqueous reactions may also remove atmospheric oxygen, especially if the planet supports a surface ocean, and these reactions are also important for our understanding of abiotic $O_2$ generation.   For example, the rate of reaction of dissolved CO and $O_2$ to reform $CO_2$, and thereby draw down atmospheric oxygen, is poorly understood, but is crucial to understanding the final balance of $O_2$ in the atmosphere (Harman et al., 2015). Similarly, weathering of surface crust (e.g. Anbar et al., 2007), and the sequestration of $O_2$ into the planetary mantle (e.g. Hamano et al., 2013; Schaefer et al., 2016) are key processes that control $O_2$ draw down, and could result in abiotic $O_2$ buildup if weak.   Conversely, if these processes are aggressive enough, they could instead result in a false negative for biologically produced $O_2$ by scrubbing photosynthetically-generated $O_2$ from a planetary atmosphere, as likely happened over the Earth's history.

Several groups have identified potential photochemical mechanisms to generate abiotic $O_2$ and $O_3$ on terrestrial planets in the habitable zone (Domagal-Goldman and Meadows, 2010; Hu et al., 2012; Domagal-Goldman et al., 2014; Tian et al., 2014; Gao et al., 2015; Harman et al., 2015).

However, the amounts of abiotic $O_2$ and $O_3$ generated in these simulations differ, in part due to different assumptions about the abundance of available catalysts to drive recombination of $CO_2$, or destruction of $O_2$ and $O_3$, which can be affected by the availability of $H_2O$, the spectrum of the star, as well as the efficiency of the CO and $O_2$ reaction in seawater.  The mechanism that could potentially produce the largest signal from $CO_2$ photolysis requires a desiccated, cold, H-poor atmosphere that inhibits $CO_2$ recombination, due to the lack of photolytic generation of the OH catalyst in the water-poor atmosphere (Gao et al., 2015).  In this scenario a catalytic cycle feedback with ozone formation may result in stable atmospheric fractions of $O_2$ near 15%.  In another proposed mechanism, recombination of photolyzed $CO_2$ can be slowed by a parent star (typically an M dwarf) with a higher FUV ($\lambda$ < 200nm) to MUV and NUV (200nm < $\lambda$ < 440nm) ratio when compared to the Sun.  The higher FUV photolyzes $CO_2$, but the lower MUV-NUV radiation inhibits the photolysis of water and other $HO_x$ chemistry that would drive recombination (Tian et al., 2014; Harman et al., 2015).  For the cases considered for this mechanism, $O_2$ abundances as high as 0.2% to 6% are predicted, with higher values corresponding to little or no $O_2$ sinks in the planetary environment. When more realistic sinks are included, abiotic $O_2$ abundances are reduced by many orders of magnitude (e.g. Harman et al., 2015).   As another possible abiotic mechanism for $O_2$ buildup, Léger et al. (2011) discussed the efficacy of known catalysts for abiotic photogeneration of $O_2$ from water-splitting on a planetary surface, but concluded that it was highly unlikely that sufficient catalysis could occur on a habitable planet to generate a false positive.  However, a subsequent study argued that this may be possible for a planet with shallow oceans, strong NUV and significantly large areas of surface TiO to produce Earth-like quantities of $O_2$ from the splitting of water over a billion years (Narita et al., 2015).

$O_3$ can often serve as a proxy for $O_2$, especially in the mid-infrared where $O_2$ does not have spectral features (e.g. DesMarais et al., 2002).  However, abiotic $O_3$ can also be formed via photolysis of the abiotic $O_2$ generated by ocean loss, with calculated values ranging from 1% to comparable to Earth's current ozone abundance for planets orbiting in the HZ of M dwarfs, depending on whether liquid water remains (Meadows et al., 2017).  However, the spectral slope of the UV radiation of the star could also photochemically generate $O_3$, even without buildup of $O_2$.   FUV radiation can favor the generation of ozone via photolysis of $CO_2$ (and $O_2$), whereas MUV or NUV radiation photolytically destroys $O_3$.  Consequently, abiotic $O_3$ could accumulate - even without significant generation of $O_2$ - for stars with the highest FUV to MUV ratios.  $O_3$ was produced at 10% of the Earth's current $O_3$ column abundance in the simulations of Domagal-Goldman et al., (2014) for M dwarf planets, without appreciable buildup of abiotic $O_2$.

Although the mechanisms that generate abiotic $O_2$ and $O_3$ are driven primarily by the interaction of the incident stellar spectrum with the planetary atmosphere, they can be balanced by the destruction or sequestration of $O_2$ and $O_3$ in the planetary environment.  These losses could be via photolysis, catalysis, catalytic recombination into $CO_2$, or interaction with the planetary surface and ocean, if present.  The net atmospheric accumulation of $O_2$ and $O_3$ is highly sensitive to these boundary conditions, which include weathering rates, and aqueous sinks for CO, all of which are currently poorly understood.

3.4 Summary of False Positives.

Recent research indicates that there are several mechanisms that could produce abiotic $O_2$ and $O_3$ in a planet's atmosphere, with each presenting a potential false positive to different degrees.   A summary of key components of this information is presented in Figure 2.   Two of the mechanisms allow water to enter a planet's stratosphere where it is photolyzed, and the H atoms lost to space, resulting in $O_2$ buildup in the planet's upper atmosphere.  Water entering the stratosphere is either enabled by loss of an ocean in a runaway greenhouse process (Luger & Barnes, 2015) – a mechanism that is most effective for late-type (i.e. less massive) M dwarfs - or via lack of non-condensable gases in the planetary atmosphere, which could affect planets orbiting stars of any spectral type (Wordsworth & Pierrehumbert, 2014).   The runaway mechanism could produce an $O_2$-dominated atmosphere of hundreds of bars, and the lack of non-condensable gases could potentially result in atmospheres that are ~15% $O_2$.  Earth-like quantities of $O_2$ have also been proposed to be generated by the splitting of liquid water by a surface TiO photocatalyst (Narita et al., 2015).   The other major class of processes that build up abiotic $O_2$ rely on the photolysis of $CO_2$ and circumstances that inhibit $CO_2$ recombination from CO and $O_2$ (Hu et al., 2012; Tian et al., 2014; Harman et al., 2015; Gao et al., 2015).   For photochemical production without atmospheric escape, $O_2$ abundances as high as 0.2% to 6% are predicted, with higher values corresponding to little or no $O_2$ sinks in the planetary environment. More realistic modeling of sinks can reduce these estimates by orders of magnitude (e.g. Domagal-Goldman et al., 2014; Harman et al., 2015). Finally $O_3$ may be considered a proxy for $O_2$ in a planetary atmosphere, and large abundances of abiotic $O_3$ may build up in the massive $O_2$-rich atmospheres possible after ocean loss (Meadows et al., 2017), although in these cases large amounts of $O_2$ will also be present.   Domagal-Goldman et al., (2014) were not able to generate large abundances of $O_2$ from $CO_2$ photolysis for habitable planets orbiting M dwarfs, but did produce potentially detectable $O_3$ column abundance as high as 10% of Earth's modern abundance.

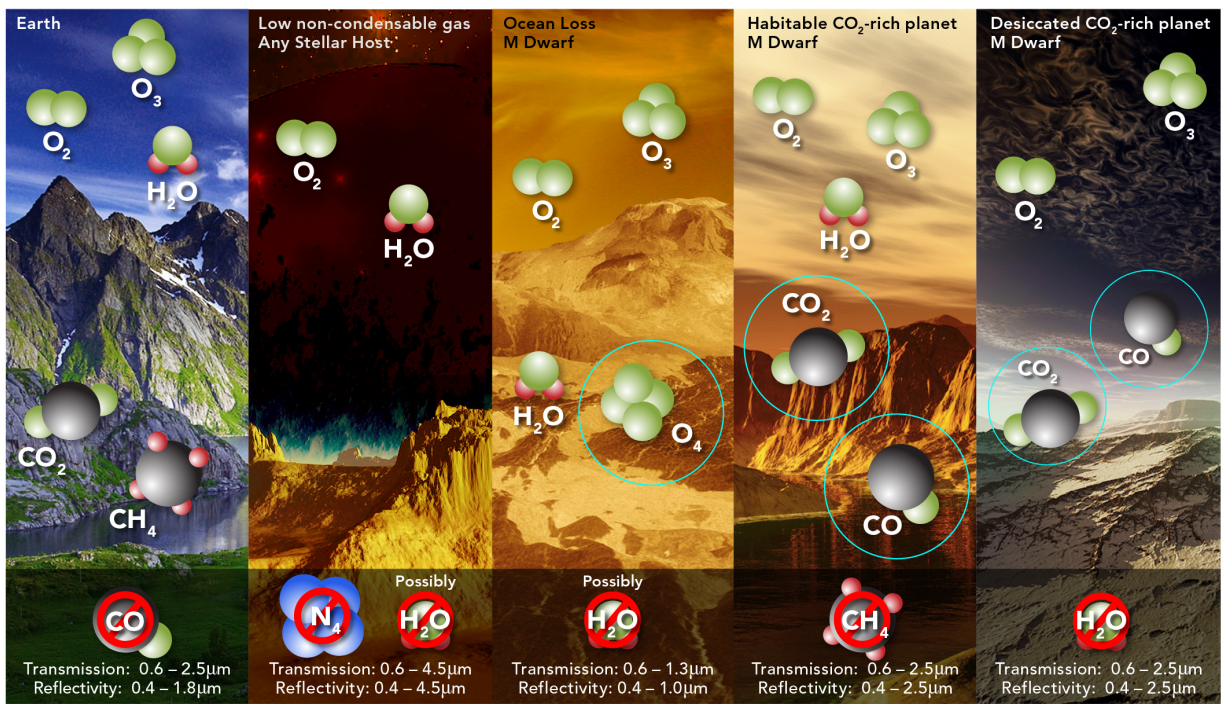

*Figure 2. Potential False Positive Mechanisms for O₂. This cartoon summarizes the atmospheric mechanisms by which O₂ could form abiotically at high abundance in a planetary atmosphere. The extreme left panel is Earth, the four panels to the right show the different mechanisms and their observational discriminants. Circled molecules, if detected, would help reveal a false positive mechanism, a lack of detection of the "forbidden" molecules in the bottom shaded bar would also help to reveal the false positive mechanism. So, for example, the presence of CO and CO₂, and the absence of CH₄ is a strong indicator for a photochemical source of O₂ from the photolysis of CO₂ on a habitable CO₂-rich M dwarf planet (Meadows 2017, Astrobiology. Figure Credit Ron Hasler).*

In most cases, however, the mechanism for abiotic production of $O_2$ or $O_3$ leaves a "tell", an impact on the planetary environment that may be detectable. These indications can range from the presence of collisionally-induced absorption from $O_2$ molecules that collide more frequently in dense, $O_2$-rich post-ocean-loss atmospheres (Schwieterman et al., 2016), CO from the photolysis of $CO_2$ (Schwieterman et al., 2016), lack of water vapor, (Gao et al., 2016), lack of collisionally-induced absorption from $N_2$ (Schwieterman et al., 2015), and the absence of reducing gases (Domagal-Goldman et al., 2014). In the following section we will describe the observations needed to search for $O_2$ in a terrestrial planetary atmosphere and to discriminate whether that $O_2$ is abiotic or biological in origin based on characteristics of the parent star and the planetary environment.

## 4.0 Observational Requirements for Detecting and Discriminating O₂ From Potential False Positives.

Three main observational techniques are likely to be able to obtain information on the characteristics of terrestrial exoplanets in the next two decades. These are transmission spectroscopy, high-contrast direct imaging, and high-resolution spectroscopy, which may also be coupled to direct imaging adaptive optics or starlight suppression systems.

Transmission spectroscopy with JWST, which will be launched in late 2018, will likely be our first opportunity to search for $O_2$ in the atmosphere of 1-3 habitable zone, Earth-sized planets (Cowan et al., 2015). Transmission spectroscopy, where a spectrum of the atmosphere is taken when a planet transits in front of and is "backlit" by its parent star, offers advantages and disadvantages for planet characterization. This technique is most likely to be used for planets orbiting M dwarfs. The increased, glancing, path length through the planetary atmosphere enhances the impact of both radiatively active gases high in the atmosphere and high-altitude aerosols on the spectrum (Fortney, 2005). Because the atmosphere is "backlit" by the star, the decrease in signal-to-noise ratio (S/N or SNR) with increasing wavelength is less problematic than in reflected light observations. However, a number of processes can limit the ability of transmission observations to probe a planet's troposphere. These processes include refraction, condensates/aerosols and the optical depth provided by atmospheric constituents. Transmission cannot observe the planetary surface, and the accessible regions of the atmosphere of a habitable planet will likely be water-poor layers above the water-preserving cold-trap, making it difficult to detect atmospheric water vapor using this technique. However, several $O_2$, $O_3$ and $O_4$ bands may be accessible with this technique.

Direct imaging has high-enough spatial resolution to separate the planet from its parent star, and also suppresses the light from that star. This allows the much fainter planet to be studied via either photometry or spectroscopy. Direct imaging can potentially probe the entire atmospheric column and even retrieve information on the planetary surface for clear-sky scenes. This technique is therefore less sensitive to aerosols, and more sensitive to the deeper, near-surface atmosphere than transmission spectroscopy (Arney et al., 2016; 2017). In the more distant future, direct imaging mission concepts such as the Habitable Exoplanet Imaging Mission (HabEx) and the Large UV Optical Infrared Surveyor (LUVOIR), one of which could be launched in the mid-2030s, would allow observations of smaller Earth-sized targets and searches for $O_2$ and other atmospheric constituents in their spectra (Bolcar et al., 2015; Dalcanton et al., 2015; Mennesson et al., 2016; Postman et al., 2010; Rauscher et al., 2015; Seager et al., 2014; Stapelfeldt et al., 2014). These missions will be enabled by starlight suppression technologies such as external starshades and/or internal coronagraphs. In addition to these space-based telescopes, a series of ground-based observatories (Extremely Large Telescopes, ELTs; e.g. Kasper et al., 2008) may be able to employ coronagraphy and adaptive optics techniques to measure spectra of terrestrial exoplanets, for a dozen targets in the habitable zones of M dwarfs. For these observations, it may be possible to use a high spectral resolution template-matching technique to disentangle exoplanet spectral lines from Earth's own atmospheric absorption features (e.g. Snellen et al., 2015). Future plans for the 30-40m diameter ground-based ELTs which will come on line in the late 2020s include enabling direct imaging observations at 10 μm for planets orbiting in the habitable zones of F, G, K stars (Brandl et al., 2012; Snellen et al., 2013; Quanz et al., 2015).

Direct imaging observations are constrained by inner- and outer-working angles (IWA and OWA) - the smallest and largest planet-star separations over which the observatory is capable of adequate starlight suppression. For coronagraphs, these angles scale with $N\lambda/D$ where N is a constant (typically of order $10^0$ for the IWA and of order $10^1$ for the OWA), $\lambda$ is the wavelength, and D is the telescope diameter. For segmented apertures with non-circular edges (e.g. the hexagonal edges of JWST), the relevant diameter for these calculations is usually that of the largest inscribed circle within the segmented aperture. For starshades, the IWA depends on the design of the starshade and the starshade-telescope separation distance. Because it is independent from the telescope aperture the starshade allows observations of close-in planets with relatively smaller telescopes apertures. However, the telescope must still be at least large enough to have the requisite spatial resolution to separate the planet and the star.

The IWA will tend to make observations of planets in the habitable zone of M dwarfs difficult, because these stars are intrinsically faint, and the HZ is 20 times closer to the star than it would be for a G dwarf. For instance, the Earth-sized planet orbiting in the habitable zone of Proxima Centauri is only 0.0485 AU away from its parent star (Anglada-Escudé et al., 2016), which corresponds to a planet-star angular separation of 37 milliarcseconds at maximum elongation. Therefore, while direct imaging of Earthlike planets orbiting the closest M dwarfs may be possible for larger observatories like the LUVOIR concept, and LUVOIR may have as much as 8% of its accessible targets orbiting M dwarfs, direct imaging missions tend to focus primarily on planets orbiting F, G, and K stars for which the planet-star angular separations are larger.

## 4.1 Target Selection

Careful target selection may help to reduce the likelihood of false positive O$_2$, although the stellar and planetary properties most likely to result in false positive O$_2$ are still being explored. Several of the proposed false positive mechanisms for O$_2$ production are more likely to occur for planets around later type M dwarfs, which may be more susceptible to early ocean loss and buildup of O$_2$ from subsequent H loss (Baraffe et al., 1990; Luger & Barnes, 2015). In contrast, planets in the habitable zones of F, G, K stars will not experience a super-luminous host star for a significant period of time. M dwarfs may also have NUV/FUV ratios that are more favorable for photochemical production of O$_2$ or O$_3$ (Tian et al., 2014; Domagal-Goldman et al., 2014; Harman et al., 2015). Given the late rise of photosynthetic O$_2$ in Earth's atmosphere, and the possibility that abiotic O$_2$ generated during the star's superluminous pre-main sequence phase may be sequestered in the planetary environment or lost to space over time (Schaefer et al., 2016), then systems older than a few Gyr may be preferred. This will allow time for biologically-generated O$_2$ to overwhelm sinks and rise in the atmosphere, and may also increase the probability that abiotic O$_2$ has been depleted.

If searching for biosignatures, it would therefore be advantageous to select older F, G, K or earlier type M dwarf targets (M0-M3) where the habitable zone is likely safer from pre-main sequence ocean loss (Luger & Barnes, 2015). It will also be important to have measurements, or be able to use proxies, to estimate the UV spectrum of the M dwarf host star. However, some of these mechanisms are more independent of the host star, and are instead tied to planetary properties such as a lack of non-condensable gas species in the planetary atmosphere (Wordsworth & Pierrehumbert, 2014) or an abundance of surface TiO (Narita et al., 2015). In these cases, abiotic O$_2$ production could occur for planets with suitable characteristics orbiting stars of any spectral type. Target selection for detailed follow up will instead rely on characterization of the planet's environment, and specifically a census of bulk gases such as N$_2$, O$_2$ and CO$_2$. More theoretical work is needed to better understand the factors that affect habitability, so that the planets most likely to be able to support life are chosen for detailed follow up.  We are also now entering a new era where the first observations of terrestrial exoplanets are being taken (e.g. de Wit et al., 2016), potentially providing empirical information relevant to habitability and false positive processes. Early observations of high-insolation exoplanets such as GJ1132b (Berta-Thompson et al., 2015) and the inner TRAPPIST-1 planets (Gillon et al., 2016; 2017) may reveal key characteristics of their planetary environments that shape abiotic $O_2$ generation and persistence, and even provide an observational test for the limits of the habitable zone.

## 4.2 Key Environmental Characteristics That Provide Context for $O_2$ Detection.

Should a promising candidate planet be identified, there are several key observations required to help place detection of atmospheric $O_2$ into context, and help discriminate between biological and abiological sources.   The first of these is detailed knowledge of the parent star, including its spectral type, age, observed or inferred UV to infrared spectrum, and a measure of its activity levels, including the amplitude and frequency of stellar flares.   The spectral type and age helps to identify the likely stellar evolutionary path of the star, which drives planetary evolution, including atmospheric loss processes, which may produce false positives.   The spectrum of the star is needed as input to atmospheric climate and photochemistry models, which will be used to determine atmospheric temperature, and the potential efficiency of photochemical production and destruction of $O_2$ and $O_3$.

Basic properties of the planet are also required, such as size, orbit, and mass.  Size is most readily determined for transiting exoplanets where the fractional dimming of the star during the transit is a direct consequence of the planet's size.  However, only a small fraction of habitable zone planets will transit (from 0.5% to 1.5% for G to M dwarf host stars).  For non-transiting planets, size is more challenging, as a size-albedo degeneracy exists in reflected light.  The range of plausible albedos for a terrestrial planet could easily span 0.06 to 0.9, potentially a factor of 7 uncertainty in the estimated size of the planet.  Observations at Mid-infrared (MIR) wavelengths – where emissivities for most terrestrial planetary surfaces are close to 1 – may be used to infer planetary size, especially if a color temperature can be derived from the shape of the MIR spectrum.   JWST MIR observations may have the ability to observe orbital phase curves for terrestrial planets in M dwarf habitable zones, while future ground-based Extremely Large Telescopes may have the ability to image terrestrial planets in the habitable zones of a small sample of M dwarfs at wavelengths near 10 μm.  If the planet can be detected in direct images at multiple epochs around its orbit, then the relative astrometry of the star and planet can be used to derive the orbital parameters.  This can be done in either reflected light or in thermal emission, but the better angular resolution available at shorter wavelengths favors the use of reflected light imaging.  If the planet is not imaged then its orbit must be determined by measuring its dynamical influence on other objects in the system.  In general this requires measurements of stellar reflex motion.  Sufficiently precise stellar astrometry can determine both the orbit and the planetary mass. Stellar radial velocity measurements can determine all the orbital elements except for inclination, and thus provide only a lower limit on the planetary mass.  The inclination ambiguity in RV orbits can be resolved by combining RV with imaging detections at more than one epoch and spread over a significant fraction of the planet's orbit.  Another means of breaking the inclination ambiguity may be via observations of emitted or reflected light from the planet in high-resolution spectroscopy. The spectral shift in atomic or molecular lines from gases in the planetary atmosphere reveal the radial velocity of the planet itself (e.g. Snellen et al., 2015), and since the planetary orbital velocity is known from the RV period, the inclination can be derived. Terrestrial planets will be extremely challenging to observe this way, and must await second or third generation instruments on ground-based ELTs. However, searching for emission from visible-light aurorae on planets orbiting M dwarfs, may help improve the contrast for planetary RV measurements (Luger et al., 2017) In the special case of transiting planets in multi-planet systems, the perturbations of one planet on another can sometimes be measured as transit timing variations and allow the planetary masses to be derived (e.g. Gillon et al., 2017).

Another key planetary characteristic needed to interpret an observation of O₂ in a planetary atmosphere is the presence of water, which could be in the form of atmospheric gas, condensed clouds, and/or a surface ocean. Although the possibility that the planet can support liquid water is more likely for an Earth-like planet within the surface liquid water habitable zone (Kasting et al., 1993; Kopparapu et al., 2013), it is not guaranteed, as water may not have been delivered to the planet during its formation (Raymond et al., 2007; Lissauer et al., 2007), or it may have been subsequently lost (Luger and Barnes, 2015; Ribas et al., 2016; Airapetian et al., 2016; Barnes et al., 2017). If water is present, it can help to rule out several of the false positive mechanisms that rely on water loss or photochemistry in an extremely water-poor atmosphere, such as those proposed by Luger & Barnes (2015) and Gao et al., (2015). Knowledge of the water vapor profile – and the detection of condensate clouds that form at the water vapor cold trap – would rule out another false positive mechanism associated with high-altitude photolysis in atmospheres without a cold trap (Wordsworth & Pierrehumbert, 2014). However, if residual water persists after water loss from either early ocean loss or lack of non-condensable gases, then abiotic oxygen and water vapor may both be seen in the spectrum (Meadows et al., 2017). Absorption features for water occur throughout the visible (0.65, 0.7, 0.73, 0.8, 0.95 μm), near-infrared (1.1, 1.4, and 1.8-2.0 μm) and MIR (6.3 μm and 18-20 μm) (Rothman et al., 2013).

Observing liquid surface water itself may be challenging and will likely only be possible for reflected light observations with the required geometry. Observing liquid surface water is a less ambiguous habitability indicator than observations of gaseous water spectral features, which could be present in uninhabitable steam atmospheres. One such method to detect liquid surface water is ocean glint, caused by specular reflection off a smooth surface (Cox and Munk 1954). Glint has been used to detect the presence of hydrocarbon seas on the surface of Saturn's moon Titan (Stephan et al., 2010). Previously, Robinson et al., (2010; 2014) used a sophisticated 3D spectral Earth model to show that the spectral behavior of Earth deviates strongly from Lambertian (i.e. isotropic) scattering behavior towards crescent phases when specular reflection would be most apparent. Forward scattering from glint may mimic forward scattering behavior of clouds, but Robinson et al., (2010) showed that ocean glint even in the presence of realistic clouds (covering ~50% of the planet) is up to a factor of two brighter than forward scattering from clouds and a non-glinting (i.e., Lambertian) ocean. Glint is most readily observed at phase angles within about 60°of new phase and for inclinations that are no more than 30° from edge on. Glint from the Earth's surface is most detectable near 0.8-0.9 μm where Earth's atmosphere is relatively transparent (although it may also be observed between the longer 1.1, 1.4, and 1.9 μm water vapor bands). Such observations may be within the reach of future direct imaging telescopes but are impossible to perform in transit transmission. In principle, phase curve observations might be able to detect the glint signal, but in practice, this is outside the realm of feasibility for JWST,. Even for the planet orbiting the closest star (Proxima Centauri b; Anglada-Escudé et al. 2016), measurement precision for detecting glint in a phase curve would need to be smaller than $10^{-8}$ (Meadows et al. 2017), but this is significantly smaller than typical JWST noise floor estimates on the order of $10^{-5}$ (Greene et al. 2016). However, this may be achievable for future ELTs and for some HZ targets observed by large space-based planet characterization missions.

Tectonic activity is critical to the long-term habitability of planets, as it both recycles elements required for life and serves as a global thermostat that regulates climate over geological timescales. Direct observations of tectonic activity will be extremely challenging, but observable impacts of tectonics on the planetary environment may be more accessible. A heterogeneous distribution of surfaces seen in maps generated using time-resolved photometry (Cowan et al., 2009) may suggest the presence of continents. Gases released from volcanic eruptions could also be sought (Kaltenegger et al., 2009; 2010), but the resulting formation and accumulation of aerosols in the upper atmosphere would likely provide larger, time-variable signals (Misra et al., 2015; Hu et al., 2013). The "ingredients" for tectonic activity could also be inferred based on the elemental composition of the system. Planetary mass and radius, informed by the Fe, Si and Mg abundance of the host star could be used to constrain the interior composition and structure and the likelihood that a planet is tectonically active (Young et al., 2014; Dorn et al., 2015)

Additional contextual information to guide interpretation of a planet's habitability can come from atmospheric pressure. Rayleigh scattering could be used to indicate pressure above the visible surface (Benneke and Seager, 2012; von Paris et al., 2013), whether that be the planetary surface or a cloud or haze layer, as is the case for Venus, although this can be compromised by a surface that strongly absorbs or scatters in the blue, as is the case for Mars and Earth, respectively. Rayleigh scattering will be less useful for planets orbiting M dwarfs, as the host star produces little radiation at the blue end of the visible spectrum where Rayleigh scattering dominates. Collision-induced features that are more sharply dependent on atmospheric pressure may provide an even more sensitive measurement of atmospheric pressure. For example $N_4$ or $O_4$ features ($N_2$-$N_2$ or $O_2$-$O_2$) seen in the near-infrared could be used to indicate higher pressures (Misra et al., 2014; Schwieterman et al., 2015b). $N_4$ absorbs strongly near 4.15 μm, just shortward of the strong 4.3 μm $CO_2$ band, and creates a strong spectral effect for abundances of $pN_2 > 0.5$ bar (Schwieterman et al., 2015b). The location of this strongly diagnostic feature in the infrared creates challenges for direct-imaging missions currently under study due to IWA and thermal background restrictions. However, it remains perhaps the only feature capable of directly confirming large quantities of $N_2$ in an atmosphere.

Finally, a census of atmospheric gases should be performed to inform planetary composition, chemistry and climate. The composition of the atmosphere will be important for assessing the potential of the planet to generate photochemical false positives, and also in calculations of the surface fluxes required to maintain $O_2$ and $O_3$ in the planetary atmosphere. It will also help identify potential sources of nutrients and energy for life at the surface. The atmospheric composition, and in particular knowledge of greenhouse gas abundances and distributions, when combined with information on the incoming stellar flux, can also be used to model the climate of the planet and its potential surface habitability.

Exoplanet characterization techniques that these types of observatories may employ include time- and wavelength-dependent photometry, phase-dependent photometry, and spectroscopy. Temporally resolved information is valuable because it can show planetary rotation and changing cloud patterns (Cowan and Strait, 2013), and longer baseline temporal measurements may reveal seasonality. Seasonal changes may be biologically-driven (e.g., Keeling et al., 1976) and so may be valuable as a separate metric from $O_2$ for detecting life on an exoplanet. Phase-dependent observations can reveal interesting features like ocean glint or the presence of forward scattering clouds, which may be water (Robinson et al., 2010). Spectra, meanwhile, open the door to detailed characterization and quantification of surface features and atmospheric gas abundances. Gas abundances derived from such spectra could be then used as inputs to atmospheric photochemical-climate models to predict the gas surface fluxes required to sustain their observed mixing ratios; quantifying these fluxes may aid in distinguishing true biosignatures from abiotic processes.

## 4.3 Detecting the O₂ signal from a Photosynthetic Biosphere

The principal means of detecting $O_2$ is via its spectral features in the visible and near-infrared at 0.69 (B-band), 0.76 (A-band), and 1.27 μm. $O_2$ is abundant and evenly mixed throughout the atmosphere on an Earth-like planet, it so is accessible to both transmission spectroscopy, which will likely probe above most of the deep atmosphere, and direct imaging, which can potentially see all the way to the planetary surface. A spectral resolution of 70 would be needed to resolve the A-band at Earth-like abundances (Robinson et al. 2016). The $O_4$ collisionally induced absorption features in the visible and near-infrared may also be used as proxies for $O_2$, and may help constrain its concentration (Misra et al., 2014). They are weakly present in the Earth's direct imaging spectrum (Tinetti et al., 2006) but become much stronger in when seen in transmission (Palle et al., 2009) or in denser $O_2$ atmospheres (Misra et al., 2014). $O_2$ is accessible to planned transmission spectroscopy and direct imaging at 0.76 and 1.27 μm, if the telescope and star will support it, but the 1.27 μm band may produce a stronger signal, especially for planets orbiting late-type M dwarfs which have low output at wavelengths shortward of 0.8 μm, and so induce low planetary transmission or reflected light signals. Ground-based high-resolution spectroscopy may be able to detect $O_2$ on some habitable zone terrestrial planets orbiting M dwarfs using the B and A-bands in the visible (Snellen et al., 2015; Lovis et al., 2016).

While obtaining well-constrained abundances for these oxygen bearing molecules may be challenging with first-generation observatories, observations of even the presence or absence of different bands of these molecules may help constrain $O_2$ abundances. The strong dependence of $O_3$ and $O_4$ features on $O_2$ concentrations could also be highly diagnostic of both chemistry and pressure. This combination of features can allow, for example, discrimination between atmospheres that contain 1% $O_2$, 20% $O_2$, and 90% $O_2$ (see Section 4.4). $O_3$ features could also point to the presence of oxygen in atmospheres with lower levels of $O_2$, when $O_2$ itself is not detectable. Figure 3 shows Proterozoic Earth with 0.01 bar of $CO_2$, 0.0003 bar of $CH_4$ and 0.1% the present atmospheric level (PAL) of $O_2$, as suggested by recent studies of chromium isotopes in geological samples from the mid-Proterozoic (~1.8-0.8 Ga; Planavsky et al., 2014). $O_2$ itself is extremely weak in the spectrum, and is not discernable in this plot, but $O_3$ produces a relatively strong UV (Hartley) band from 0.2-0.3 μm that is not saturated, unlike the present Earth's. Consequently, accurate measurement of the bottom of the band - which is needed to quantify abundance - is potentially easier than for Earth.

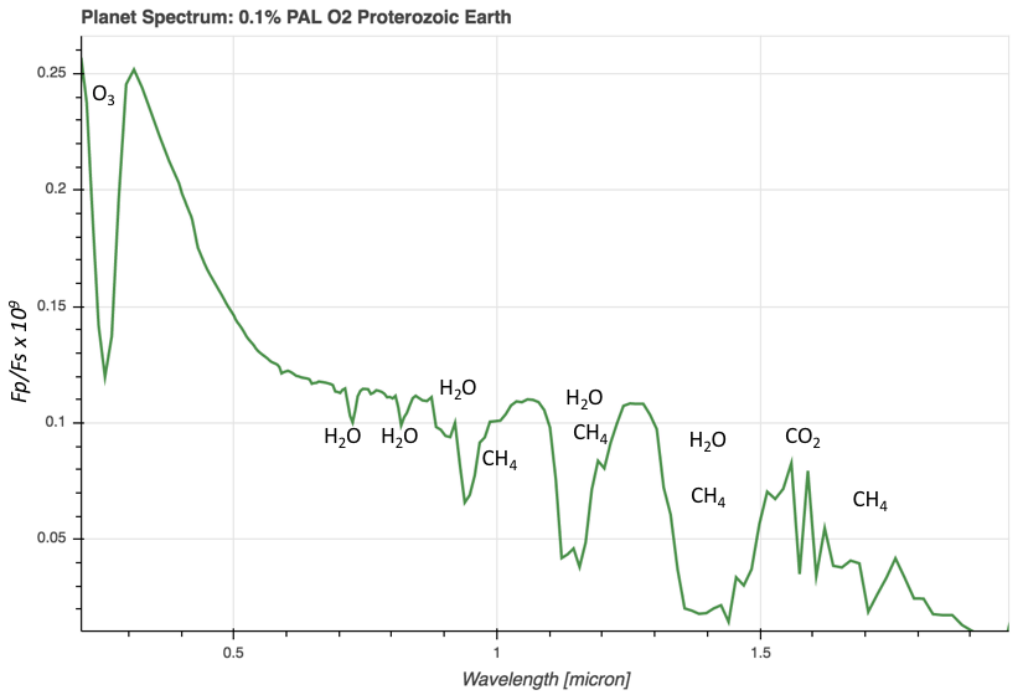

*Figure 3. Proterozoic Earth with 0.1% PAL $O_2$. Note the absence of strong $O_2$ features, but its photochemical byproduct $O_3$, produces a relatively strong feature in the UV (Credit G. Arney).*

Searching for $O_2$ in the atmosphere of Proxima Centauri b using ground-based telescopes may be possible in the next 5 years.   Perhaps the most rapid path to this goal was suggested by Lovis et al., (2016) who proposed upgrades and the development of a coupling interface between the SPHERE high-contrast imager and the new ESPRESSO high-resolution spectrograph on the 8.2m Very Large Telescope.  The combined high-contrast/high-resolution technique would provide the desired angular separation and contrast sensitivity of $10^{-7}$ in reflected light.  This would make it possible to probe the $O_2$ bands at 0.63, 0.69 and .76 μm, along with water vapor at 0.72μm and $CH_4$ at 0.75μm.  Lovis et al., (2016) calculate that a ~4-sigma detection of $O_2$ on an Earth-like Proxima Cen b could be made in about 60 nights of telescope time, spread over 3 years to observe the planet at maximum separation from the star.   Even better results for these M dwarf HZ planets are likely possible with the Extremely Large Telescopes slated for first light in the late 2020s.

Future large space telescopes should also be able to detect photosynthetic $O_2$, primarily for planets orbiting F, G, K stars. The design requirements for the HabEx and LUVOIR observatory concepts include the ability to detect $O_2$ in the atmospheres of neighboring Earth-like exoplanets (e.g. Mennesson et al. 2016, Dalcanton et al. 2015). Whether a planet's $O_2$ spectral features are in practice detectable with these observatories will depend on several parameters, including the planet's $O_2$ abundance and the distance of the planet-star system to Earth. For example, for a 6.5 m (JWST-sized) optical telescope with coronagraphic capability, if the observatory's inner-working angle was given by $2\lambda/D$, the $O_2$ A-band on Earthlike planets orbiting sunlike stars would be unobservable for planet-star systems farther than about 20 pc, and at this distance, very long integration times (> 100s of hours) would be required to obtain sufficient S/N to detect Earthlike $O_2$ levels with an observatory of this size. A larger telescope (12.7 m) could detect $O_2$ out to about 40 pc given the same IWA – although, again, very long integration times would be required to detect the $O_2$ feature for planets at this distance. Figure 4 shows the integration time as a function of wavelength centered on the $O_2$ A-band required for a 15m LUVOIR-class telescope to obtain SNR = 10 for a planet identical to modern Earth orbiting a solar twin at a distance of 10 parsecs. The spectral resolution assumed here is 150. For these assumptions, 10s of hours are required to obtain SNR=10 across most of the wavelength range shown. Figure 4 shows the spectrum for the planet that could be obtained across the UV-VIS-NIR wavelength range in 30 hours per coronagraphic bandpass for this same observatory. The significant increase in the size of the error bars at longer wavelengths is due to thermal emission from the telescope, which is assumed to be heated to 270 K. The spectral resolutions assumed for the UV-VIS-NIR channels for both Figures 4 and 5 are R = 20 for the UV ($\lambda$ < 0.4 um), R = 150 for the visible (0.4 < $\lambda$ < 0.85) and R = 100 in the NIR ($\lambda$ > 0.85). Both were simulated using a coronagraph noise model based on the one described in Robinson et al. (2016)

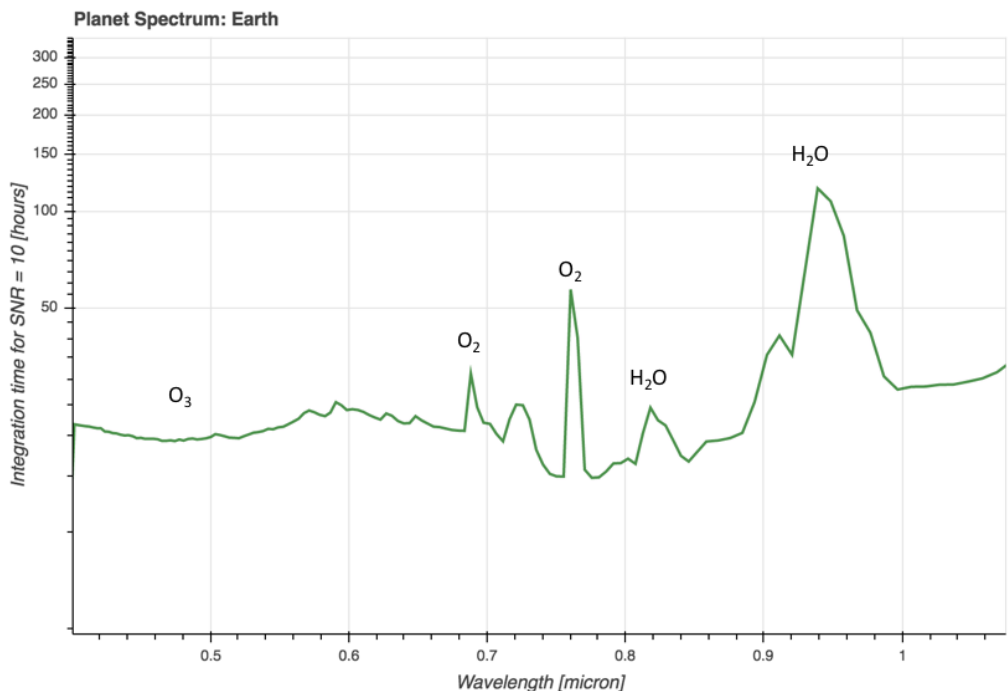

*Figure 4. The integration time as a function of wavelength required to obtain a SNR = 10 for the modern Earth orbiting a star at 10 pc for a 15 m LUVOIR-class telescope. The spectrum is roughly centered in the $O_2$ 0.76um A-band. Spectral resolution = 150 in the visible. SNR = 10 can be obtained in 10s of hours at most wavelengths (Credit G. Arney).*

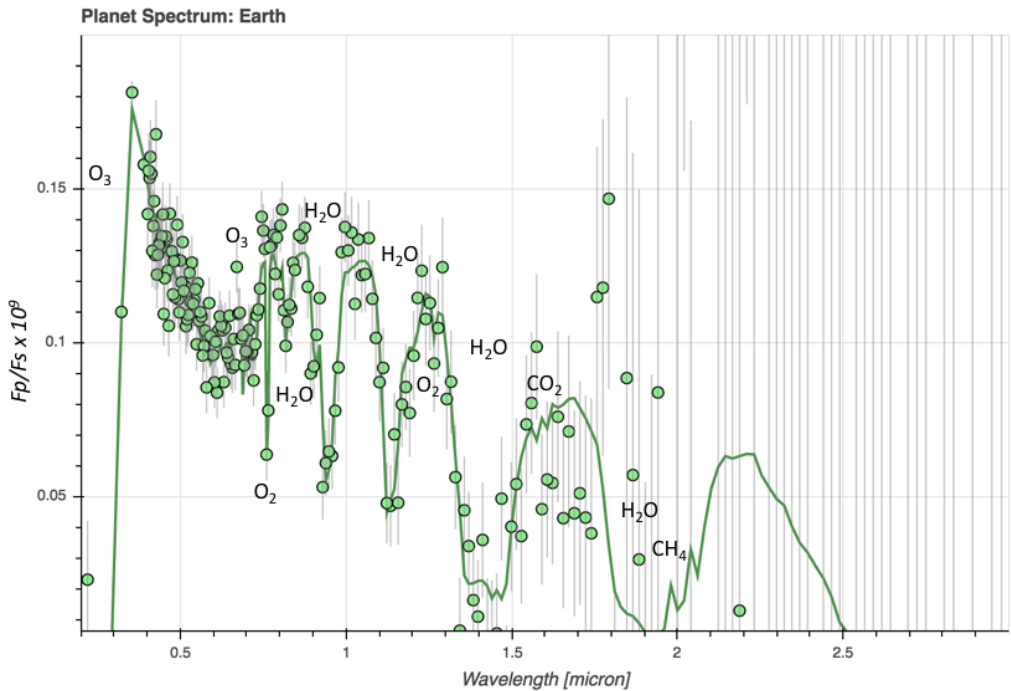

*Figure 5. The reflectance spectrum obtainable in 30 hours by a 15 m space-based telescope observing modern Earth orbiting a sunlike star at 10 pc. The grey bars denote the noise level, which increases significantly at wavelengths longward of 1.8 μm due to thermal radiation from the telescope, which is assumed to be heated to 270K (Credit G. Arney).*

## 4.4 Discriminating False Positives

Once $O_2$ is detected in a planetary spectrum, searching for additional observed environmental features to place the $O_2$ in context will aid in discriminating false positive $O_2$ from true biological $O_2$. Below we describe currently known methods for observational discriminants for abiotic and biological $O_2$.

### 4.4.1 Identifying $O_2$-Buildup from Ocean Loss

A proposed discriminant of the massive $O_2$-rich atmosphere produced by the loss of oceans of water during the pre-main sequence phase of the host star is the appearance of $O_4$ collisional complexes in the planet's spectrum (Schwieterman et al., 2016); these features are caused by collision induced $O_2$-$O_2$ absorption and are highly sensitivity to atmospheric density. $O_4$ features occur at 0.345, 0.36, 0.38, 0.445, 0.475, 0.53, 0.57, and 0.63 μm in the visible and at 1.06 and 1.27 μm in the NIR (Greenblatt et al., 1990; Hermans et al., 1999; Maté et al., 1999; Richard et al., 2012; Schwieterman et al., 2016). These strong $O_4$ features would indicate an $O_2$ atmosphere that is likely too massive to be biologically produced (Schwieterman et al., 2016). The most diagnostic spectral indicators of high $O_2$ buildup ($pO_2 > 1$ bar) in transmission observations are will likely come from 1.06 and 1.27 μm $O_4$ bands (Misra et al., 2014; Schwieterman et al., 2016) and are shown in Figure 6. These longer wavelength bands are not as badly affected by the increase in Rayleigh scattering as their counterparts at shorter wavelengths. Using the photon-limited JWST instrument model of Deming et al. (2009), Schwieterman et al. (2016) find that JWST/NIRISS could detect the 1.06 and 1.27 μm bands with a S/N of ~3 assuming 65-hour integrations (10 transits) of an Earth-like planet around a star like GJ 876 (M4V; R=0.39 $R_\odot$) (Figure 7). However, other, more conservative estimates suggest no features in any high-molecular weight atmosphere could be detected by JWST due to a systematic noise floor (e.g., Greene et al., 2015). For direct imaging observations the shorter wavelength $O_4$ bands in the visible are now prominent against the Rayleigh scattering slope (Figure 8). These bands can be obtained even within the shorter wavelength range constrained by the IWA for more distant targets, or for planets that are closer to their star as is the case for M and K habitable zone planets.

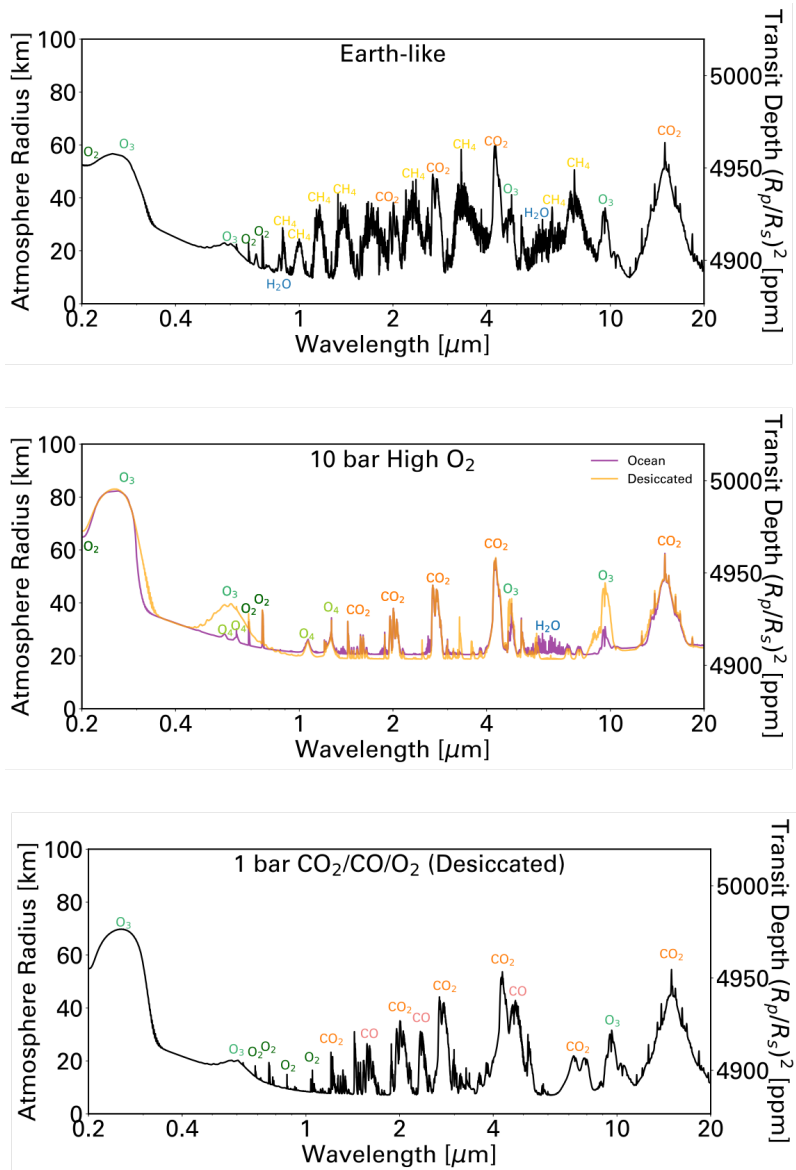

*Figure 6: Transit transmission spectra of potential planetary environments with different $O_2$ abundances for planet orbiting the M5.5V star Proxima Centauri (Meadows et al., 2017),*

*illustrating spectral features that can help distinguish photosynthetic from abiotically generated $O_2$ in a planetary atmosphere. From top to bottom: self-consistent Earth-like atmosphere with 50% cloud cover (21% $O_2$); 10 bar abiotic $O_2$ (95% $O_2$) atmosphere produced by early ocean loss with ocean remaining (purple) and desiccated (orange); 1 bar desiccated $CO_2/CO/O_2$ atmosphere which has reached a kinetic-photochemical equilibrium between the photolysis rate of $CO_2$ and kinetics-limited recombination (15 % $O_2$). Effective atmospheric radius in km are on the left y-axes and transit depth is shown on the right y-axes. The photosynthetic source for the $O_2$ in the Earth like case is made more likely by the presence of $O_2/O_3$, water, and methane. High $O_2$ cases with and without water are distinguished by the presence of O4, and the behavior of the 0.5-0.7 μm Chappuis band that is sensitive to tropospheric $O_3$, which is more abundant in the desiccated case. The desiccated chemical equilibrium atmosphere is easily distinguished by its high levels of CO.*

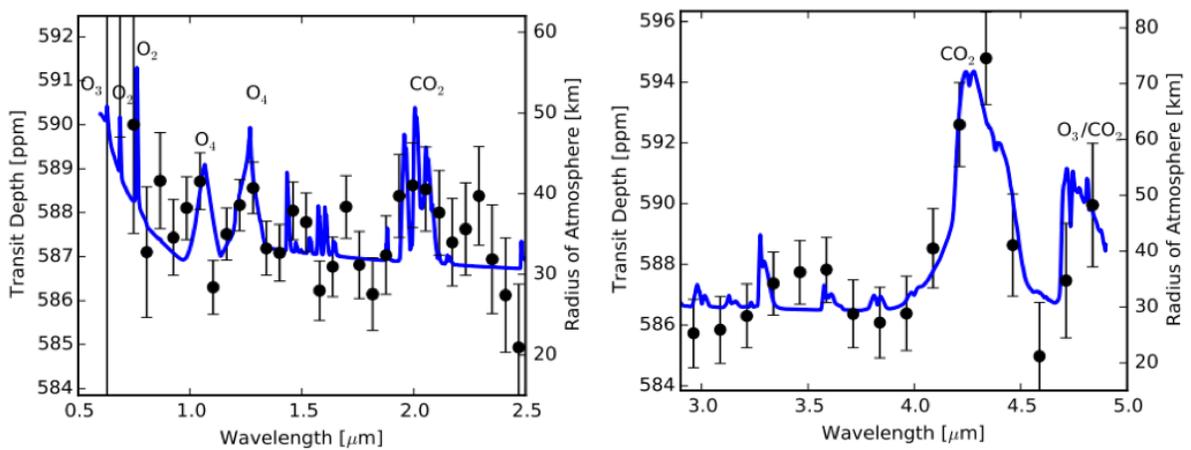

*Figure 7: Synthetic transmission spectrum of high-$O_2$ atmosphere with NIR $O_4$ features at 1.06 and 1.27 μm. The model atmosphere is a hypothetical 100 bar $O_2$ atmosphere left behind by massive H-escape during pre-main sequence evolution (Luger and Barnes, 2015). Data and error bars (1σ) for simulated JWST-NIRISS (left) and JWST-NIRSpec (right) were calculated with the noise model of Deming et al. (2009) assuming 65 hour integrations (10 transits of an Earth-size planet around GJ876) and photon-limited noise. Figure adapted from Schwieterman et al. (2016).*

### 4.4.2 Identifying Abiotic $O_2/O_3$ from $CO_2$ Photolysis

An atmosphere with significant $O_2$ produced by $CO_2$ photolysis is likely to also have three observable features: 1) a sufficient UV flux from the host star; 2) high (> 0.1 bars) $CO_2$ in the atmosphere (Domagal-Goldman et al., 2014); and 3) a significant quantity of photochemically-produced CO. Therefore, strong CO features in the presence of abundant $O_2$ and $CO_2$ could signal this false positive scenario (bottom panels of Figures 6 & 8). $CO_2$ exhibits strong spectral features near

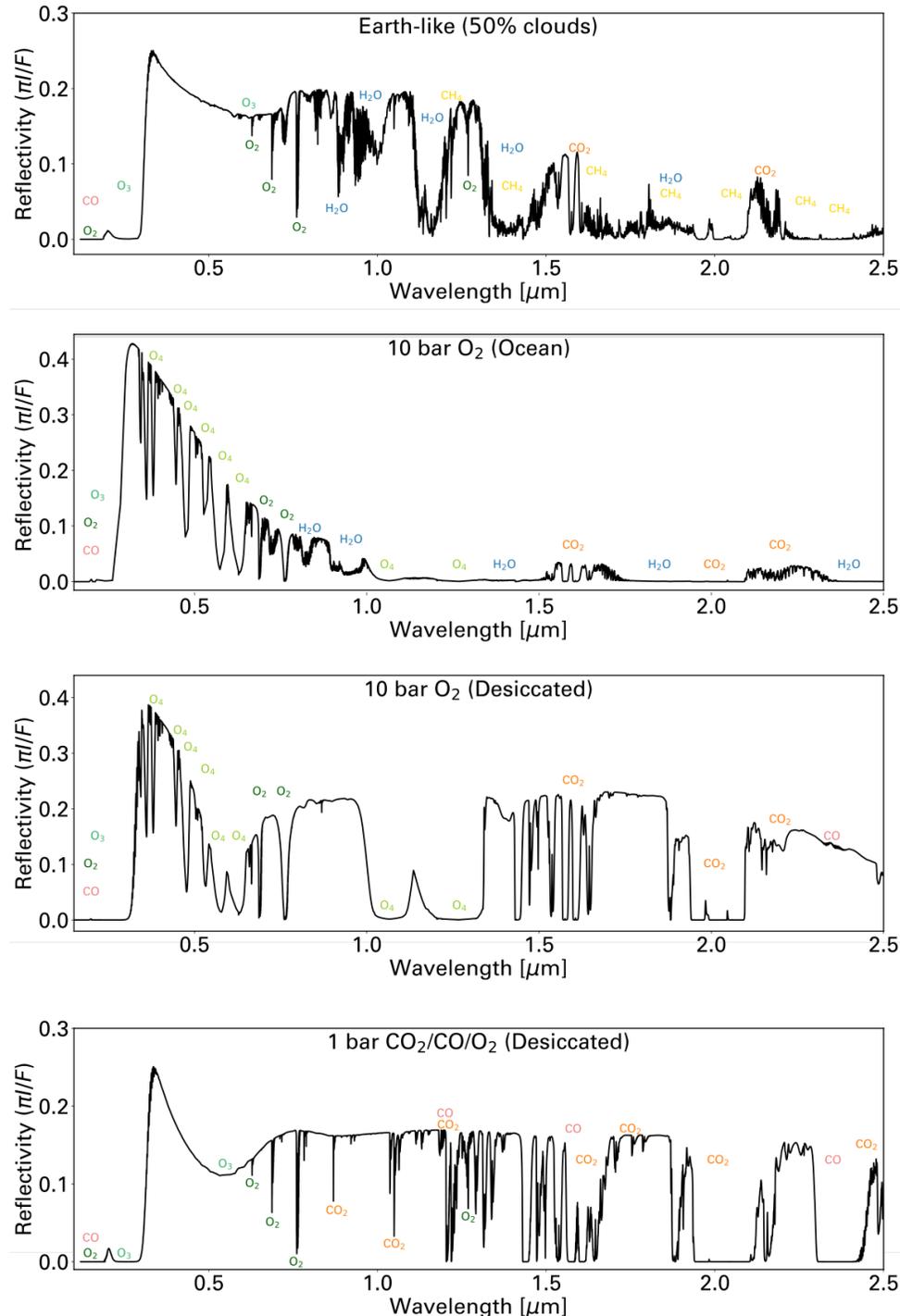

Figure 8: *Reflected light spectra of potential Proxima Centauri b climates, from top to bottom: self-consistent Earth-like atmosphere with 50% cloud cover, 10 bar abiotic $O_2$ atmosphere with ocean, 10 bar desiccated $O_2$ atmosphere, 1 bar desiccated $CO_2/CO/O_2$ atmosphere in kinetic-photochemical equilibrium (see Meadows et al, 2017). The oxygen content in these atmospheres are produced from very different mechanisms: Earth $O_2$ is biological, the 10 bar abiotic oxygen*

*atmospheres result from the super-luminous pre-main-sequence evolution of the planet's M dwarf host star, and the last atmosphere is a 1 bar $CO_2$ atmosphere that has < 1 ppm hydrogen, resulting in a slow $CO_2$ recombination timescale compared to its photolysis rate and Earth-like levels of oxygen. However, because hydrogen is also required destroy ozone, this atmosphere exhibits more ozone and a stronger Chappuis ozone band at ~0.6 μm. The 10 bar $O_2$ atmospheres are easily distinguished from Earth-like atmospheres by their deep, wide $O_2$-$O_2$ ($O_4$) collision-induced absorption bands. Moist vs desiccated abiotic $O_2$ cases are distinguished primarily by the presence or absence of water features. Around M dwarf stars, Earth-like planets with biological and geophysical sources of methane result in longer atmospheric lifetimes and correspondingly deep, observable methane features.*

1.65, 2, and 4.3 μm, and weaker bands occur near 0.78, 0.87, 1.05, and 1.2 μm (Rothman et al., 2013). CO absorbs strongly near 2.35 and 4.6 μm. In practice, for exoplanet direct imaging space observatories that are not cryogenically cooled to lower than room temperature, simulations show that wavelengths longer than about 1.8 μm may be effectively unobservable, except for the brightest targets, due to the overwhelming thermal background of the telescope itself (Figure 5). In this case, only the weaker 1.65 μm $CO_2$ band and none of the CO bands would be available. Additionally, the NIR spectrograph channel may not extend to long enough wavelengths to access these features. This would make it challenging to identify this false positive in direct imaging. However, these longer wavelengths may be accessible to JWST. Assuming the 'worst case' scenario for abiotic $O_2$ generation ($pO_2 = 0.06$ bars) for an Earth-like M-dwarf HZ planet with a prebiotic $N_2$-$CO_2$-$H_2O$ atmosphere (Harman et al. 2015), CO and $CO_2$ bands shortward of 4.6 μm could be detected with an S/N ~ 3 assuming a 65-hour (10 transit) observations by JWST NIRISS (0.6-2.9 μm) and NIRSPEC (2.9-5.0 μm), assuming photon-limited noise (Figure 9; Schwieterman et al., 2016). As above, the $CO_2$ and CO bands are the strongest features in the simulated spectrum.

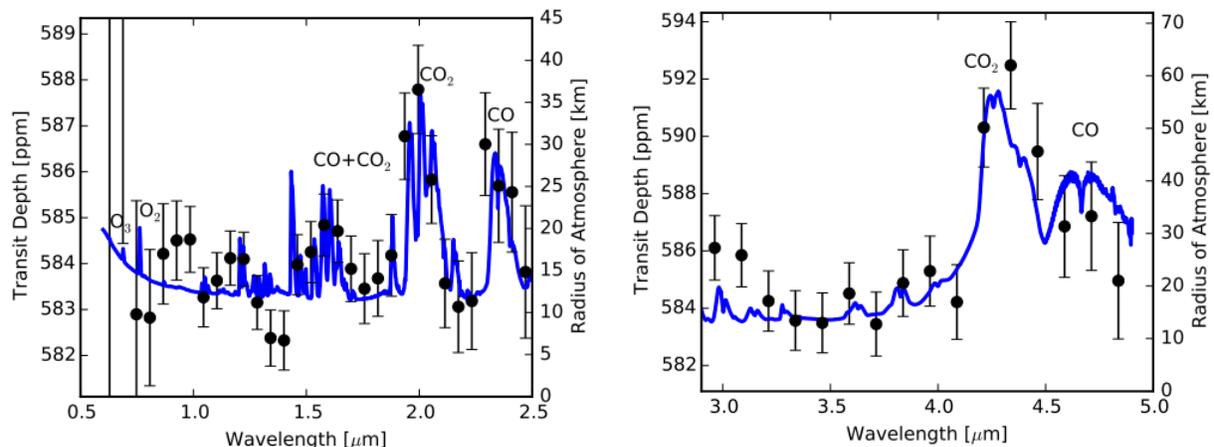

*Figure 9: Synthetic transmission spectrum with photochemical $O_2$ and CO features. Model prebiotic atmosphere in photochemical equilibrium around GJ876 from Harman et al. (2015) containing ~6% abiotic $O_2$ and ~1% CO. Data and error bars (1σ) for simulated JWST-NIRISS (left) and JWST-NIRSpec (right) were calculated with the noise model of Deming et al. (2009) assuming 65 hour integrations (10 transits of an Earth-size planet around GJ876) and photon-limited noise. Figure adapted from Schwieterman et al. (2016).*

### 4.4.3 Constraining Abiotic O₂ from low N₂ inventories

The abiotic $O_2$ mechanism posited by Wordsworth & Pierrehumbert (2014) suggests that low non-condensing gas inventories on Earth-like planets could lead to abiotic $O_2$ buildup by lifting the tropospheric cold trap, allowing water to reach the stratosphere where it could be photolyzed. The H would escape, leaving the O behind to potentially build up in the atmosphere as $O_2$. However, this scenario could be ruled out if the non-condensing gas abundance could be directly constrained. Throughout Earth's long 4.5-billion-year history, the dominant non-condensable gas has been $N_2$. The $N_4$ ($N_2$-$N_2$ CIA) feature overlaps with the 4.3 μm $CO_2$ band (Figure 8), but absorbs over a much broader region (Lafferty et al., 1996; Schwieterman et al., 2015b). In reflected light spectra, the influence of $N_4$ on the 4.1 μm region is strongly dependent on $N_2$ abundance beyond $P_0$=0.5 bars (see Figure 9; Schwieterman et al. 2015b). There is also a much weaker harmonic $N_4$ band at 2.1 μm (Shapiro & Gush 1966), seen in the reflectance spectra of nitrogen ices on outer solar system bodies (Grundy & Fink 1991), but this band is currently poorly characterized and probably too weak to create a measurable impact on plausible planetary atmospheres.

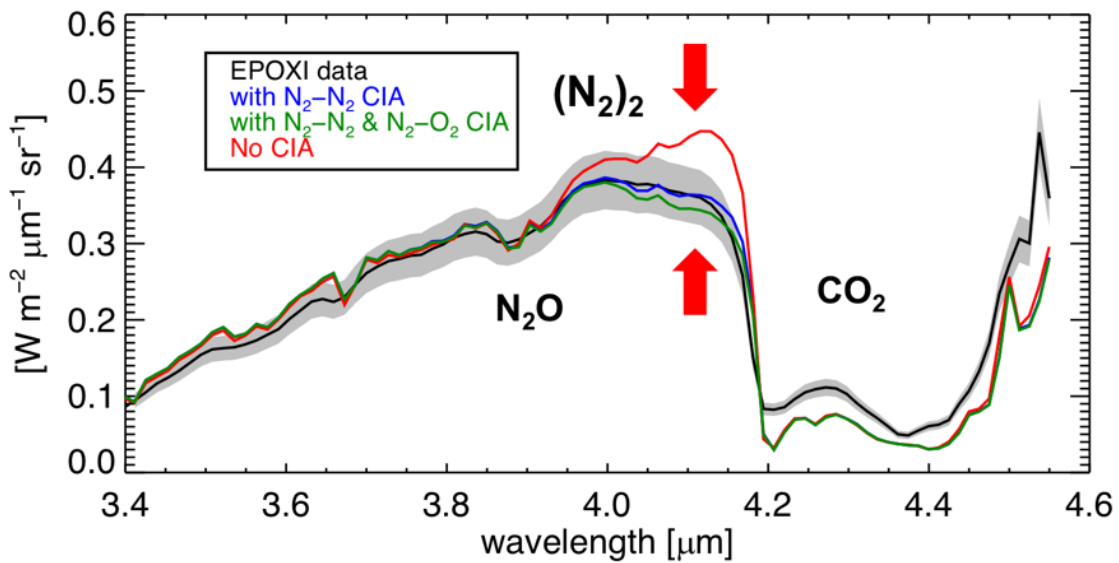

*Figure 10: The impact of $N_4$ absorption in Earth's disk-averaged NIR spectrum. This plot shows the spectral differences between the cases with $N_4$ absorption (blue & green) and the case without it (red), when compared to the EPOXI observations of the Earth in this wavelength range (black). $N_4$ is clearly required to match the observed spectrum of the Earth near 4.1μm. The gray band shows the calibration uncertainty for the EPOXI data (Klassen et al., 2008). Figure adapted from Schwieterman et al. (2015b).*

The spectral transmission depths due to $N_4$ will vary depending on the abundance of $CO_2$ and the overall mean molecular weight of the atmosphere, but can reach 10 ppm for an $N_2$-dominated atmosphere for an Earth-size planet transiting an M5V (R=0.2 $R_\odot$) star with larger transmission depths possible for atmospheres with a low molecular weight component such as $H_2$ (Schwieterman et al., 2015b).

## 5.0 An Integrated Observing Strategy for Finding and Identifying an Oxygenic Photosynthetic Biosphere.

The broad steps in a potential observing strategy to identify $O_2$-bearing terrestrial planets include detecting the planet in the presence of astronomical noise sources, preliminary characterization of the planet, detection of $O_2$, the search for false positives, and using the combination of planetary environmental characteristics and the results of the false positives searches to discriminate between biotic and abiotic sources for $O_2$. These steps for a space-based direct imaging telescope are illustrated in Figure 11. For these steps, we specifically assume that spatially-resolved spectra over the field of view are possible, and that the mission can observe in the UV-NIR wavelength range. Other observing strategies may be more appropriate for observatories without these capabilities. While this strategy will be most appropriate for a telescope concept such as LUVOIR or HabEx, a similar or complementary approach to characterizing planetary environmental context and searching for biosignatures could be developed for MIR-FIR direct imaging, or for transit spectroscopy.

**Planet Detection:** One of the first challenges will be to detect the planet and discriminate it from potential background sources, which include "exozodiacal dust" ("exozodi") from the planetary system dust disk, and background stars and galaxies. This could be done one of two ways, either with multi-epoch observations—revisits to the planetary system spread out over a longer time period—that will likely more conclusively discriminate the planet from background sources using its orbital motion. Multi-epoch observations will be spread out over a longer time period, but will simultaneously acquire the position of the planet against the background sky, and this information can be used to determine mass and orbital parameters such as semi-major axis and eccentricity (Stapelfeldt et al., 2015). An alternative, potentially faster method of planet detection would be to use a single visit with multi-wavelength imaging to look for point sources near the star. Additional information could then be used to determine if these sources are more or less likely to be planets or background objects. For example, point sources detected within the exo-zodi are more likely to be planets than sources detected outside the disk. During this detection phase, photometric colors – or low resolution, spatially-resolved spectra across the field of view, could enhance our ability to discriminate planets from astrophysical noise sources such as stars, brown dwarfs, and galaxies (Seager et al., 2015).

**Preliminary Characterization:** If the planet is detected in a single visit, then planetary photometric colors could be used to not only discriminate planets from background noise sources, but to attempt to discriminate between major types of planets. However, planetary environments are rich and complex, especially terrestrial planets which may display inhomogenous surfaces, volatiles in multiple phases and a broad diversity of atmospheric compositions. Colors would therefore provide only a possibly ambiguous initial classification (Hegde & Kaltenegger, 2013), because strong degeneracies in environment types that fit a given region of the color-color diagram likely exist (Krissansen-Totton et al., 2016). Therefore, a potentially more valuable initial method of discriminating between planetary targets would be to obtain low resolution (R~10) spectra to search for molecules with broad absorption features, such as water vapor and ozone in the visible and NIR. While neither the $H_2O$ nor the $O_3$ Chappuis band would be uniquely diagnostic of oceans or life, respectively, the detection of either of these gases would be consistent with habitability and biology. Water vapor features are present near the $O_2$ A-band at 0.72 μm and 0.85 μm, and at increasing strength at 0.95 μm, 1.14 μm and 1.4 μm.

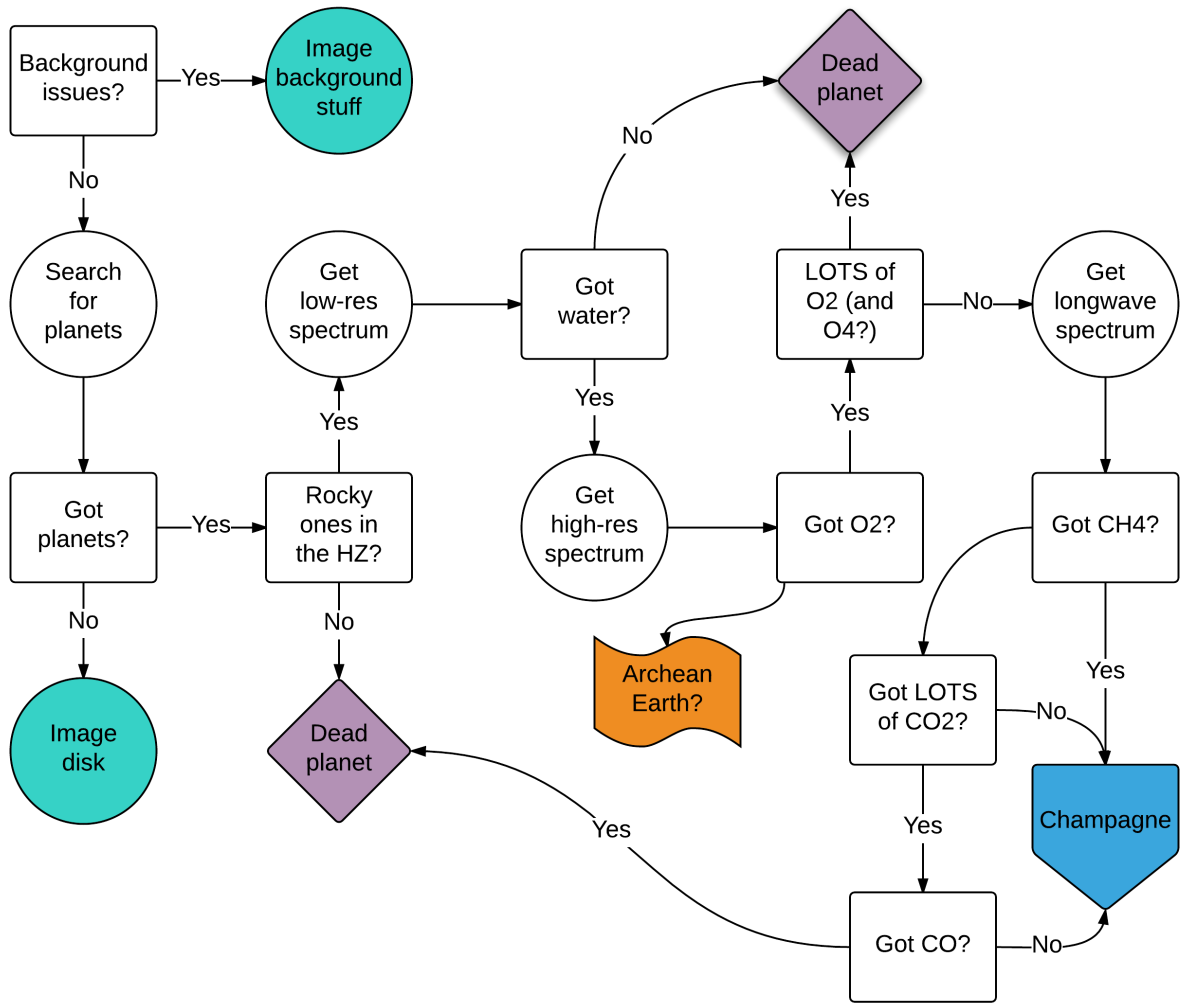

*Figure 11: A flowchart cartoon of the steps to be taken in searching for a photosynthetic biosphere on an extrasolar planet, and in interpreting any $O_2$ that is observed. (Credit: Shawn Domagal-Goldman).*

If multi-epoch photometric/astrometric observations are used for detection, then the initial characterization could include determining the planet's orbit and therefore semi-major axis and eccentricity. These parameters will help determine the planet's insolation, which will constrain climate models of the planet, and indicate whether or not the planet is in the habitable zone. Mass determination may be done by using or combining any follow-up or existing RV data or the the observed position of the planet. The orientation of the dust disk may also help break the RV degeneracy between mass and inclination of the planetary system, although an orbital solution for the planet will do this also. At this stage, direct constraints on planetary size will only be possible if the planet transits, or if ground- or space-based MIR observations are able to observe the target. Size or mass may help indicate planets that are more likely to be terrestrial (Rogers et al., 2015).

Time-resolved, multi-wavelength photometry, even in a single 20% coronagraphic bandpass, and for a single visit, could be used to map the planet and search for surface inhomogeneities that may indicate a terrestrial planet (Cowan et al., 2009; Kawahara and Fujii, 2010; Lustig-Yaeger et al., 2017). For multi-epoch observations, photometry at near-infrared wavelengths (0.8-0.9 µm, and between water bands in the 1.0-1.4µm region) comparing brightness at different points in the planet's orbit near gibbous and crescent phases could be used to search for ocean glint (Robinson et al., 2010). Glint observations may be challenging, as the crescent phase puts the planet close to the star, and so perhaps interior to the IWA, and the inclination of the planetary system to the observer must be closer to edge-on than face-on, otherwise glint will not be observed. Even so, LUVOIR-class missions with 16m mirrors may be able to search for glint for a subset of their terrestrial planets. Detection of any of these planetary characteristics, terrestrial mass or size, interesting maps, glint, or hints at the presence of water or ozone would be enough to motivate further observations at higher spectral resolutions (R ~ 150).

**The Search for a Oxygenic Photosynthetic Biosphere With High Resolution Spectroscopy:** Once a promising target has been identified, significant time could be invested in obtaining a high-resolution spectrum. At an R of ~150, individual features from many molecules can be detected. $O_2$ should be sought, and if detected, deeper observations should be taken to constrain its concentration. The oxygen A-band at 0.76 µm is a narrow but deep feature at Earthlike concentrations, and the wavelength range and spectral resolution of potential future direct imaging observatories such as HabEx and LUVOIR will be designed to detect it in the atmospheres of our exoplanetary neighbors. Detecting the $O_3$ UV Hartley band as a proxy for $O_2$ in situations where the $O_2$ concentrations is too low to product observable spectral signatures may also be possible. This may be the case for planets like the Proterozoic Earth, where its 0.1% PAL $O_2$ is likely not detectable in reflected light in the visible-NIR with currently anticipated instrumentation, but the $O_3$ Hartley band produces a deep feature for λ < 0.3 µm. As stars are fainter in the UV, longer integration times, wider bandpasses, or both, will be required to obtain high SNR observations of UV absorption features. The 30 hour observation of an Earth twin planet at 10pc with a 15 m telescope shown in Figure 6 returned adequate S/N but for a spectral resolution of 20, significantly lower than the R = 150 assumed for the visible channel. If $O_{2,}$ or $O_3$ is not detected, but $H_2O$ is, this may not indicate a lifeless planet, but rather one with a different dominant metabolism (e.g. methane-producing), or an environment that leads to a false negative, as would have been the case for the early Earth.

**Further Characterization and Elimination of False Positives:** Other molecules should be sought to characterize the planetary environment, including $O_3$, $O_4$, CO, $CO_2$ and $CH_4$, as well as clouds and aerosols. The presence of $O_4$, especially for a planet orbiting a late-type star, may indicate that any $O_2$ present is the result of ocean loss and H escape due to $H_2O$ photolysis. In direct imaging, the strong $O_4$ collision induced absorption features in the visible (0.34-0.7 µm) would likely be detectable in the same observations required to detect the $O_2$ (0.76µm), and could potentially be in the same coronagraphic bandpass. In comparison, for transmission observations, $O_4$ bands in the near-infrared at 1.06 and 1.27µm would be more detectable, as they avoid the loss of atmospheric opacity engendered by Rayleigh scattering at shorter wavelengths. Note also that ground-based high-resolution spectroscopy will not be able to observe $O_4$ collisionally induced absorption, and so this false positive discriminator is not available to this technique.

If abundant $CH_4$ is observed with $O_2$, this would make most photochemical false positives less likely, as $CH_4$ is a sink for photochemically-generated $O_2$. It is also less likely to build up to detectable levels in the massive $O_2$ atmospheres of ocean loss planets. Thus, detection of $CH_4$ in an atmosphere with $O_2$ features would rule out ocean loss and photochemical false positives, and suggest a high $O_2$ replenishment rate and a likely biogenic source of the $O_2$. However, $CH_4$ features may be challenging to observe at modern Earth-like concentrations because its strongest features are longward of 2.5 μm where telescopic thermal emission and inner working issues may be significant. Weaker bands in the visible (0.7-0.9μm) would require abundances of $CH_4$ in excess of those seen in Earth's atmosphere today, but are within the wavelength ranges anticipated for direct imaging spacecraft and ground-based high-resolution spectroscopy (Lovis et al., 2016). Indeed, as described in section 2.2, there may have been no period in the Earth's history where both $O_2$ and $CH_4$ coexisted in high abundance.

Photochemical generation of $O_2$ and/or $O_3$ from $CO_2$ photolysis (Domagal-Goldman et al., 2015; Gao et al., 2015; Harman et al., 2016), should be associated with high amounts of $CO_2$ and CO in the planetary atmosphere. $CO_2$ has a weak band near 1.6 μm, which is undetectable on modern Earth but would likely have a strong feature for atmospheres with enough $CO_2$ to produce $O_2$ or $O_3$ photochemically (Domagal-Goldman et al., 2014). $CO_2$ has a much stronger band at 4.2μm, although this latter band will be difficult to access for warm telescopes dominated by thermal radiation in this spectral region. CO also absorbs near 2.35 and 4.6 μm, where a warm mirror will also preclude observation. Additionally, these longer wavelengths will require a large telescope diameter to avoid falling within the inner working angle for observatories cold enough to see them, or an extremely large starshade to provide suppression at those longer wavelengths. Transit transmission observations, however, may allow access to longer wavelengths in the near term, compared to direct imaging techniques, and ground-based observations may also be attempted for the 2.35μm CO band.

**Detailed Planet Characterization and the Search for Secondary Biosignatures.**

If all known false positive mechanisms for the presence of $O_2$ have been eliminated, and if photochemical models and the presence of $CH_4$ suggest a high $O_2$ flux into the environment, then this would make a biogenic source more likely. This conclusion can be bolstered not only by ruling out false positives, but by searching for secondary confirmation of the dominant metabolism that is believed to have been discovered. In the case of oxygenic photosynthesis, this could come in the form of secondary biosignatures such as detection of a red-edge or other pigment-associated feature in multi-wavelength, time-resolved photometric mapping (Lustig-Yaeger, 2017) or in seasonally-varying concentrations of $O_2$ $CO_2$, or $CH_4$ (Reinhard et al., 2016; Meadows 2008), or the detection of other biosignatures such as $N_2O$ (Schwieterman et al., 2017). Eventually, other biosignatures and their false positives will be identified and sought, and the measurements obtained from any single planet will be placed in the broader context of similar measurements made on other worlds (Fujii et al., 2017; Walker et al., 2017), further informing our understanding of comparative planetology, and the prevalence of life in the Universe.

## 6.0 Discussion and Lessons Learned

The recent detailed study of $O_2$ has advanced the field of biosignatures by providing an in-depth exploration of the impact a planetary environment can have on our ability to detect and recognize biosignatures. $O_2$ can therefore serve as an exemplar for a more generalized framework for biosignature detection. It has long been considered an excellent biosignature in part because of its characteristics in the modern Earth environment. $O_2$ is an abundant gas that is stable against photolysis, and so it is evenly mixed throughout the atmosphere. This increases its detectability even in the presence of clouds, or when using techniques like transmission spectroscopy that do not probe deep into the planetary atmosphere. In addition, for the Earth, there are no known abiotic processes that would produce it in large abundance. The study of $O_2$ as a biosignature draws heavily on our understanding of how the Earth and life co-evolved over time, and the planetary processes that may have suppressed or enabled $O_2$'s rise from a trace gas to one of the principal components of our planetary atmosphere. We have learned that a biosignature will not manifest itself in a planetary atmosphere simply because life evolved, but rather as a result of the complex interactions of life with the planet and the host star, often over billions of years. The study of $O_2$ throughout Earth's history has introduced the concept of a false negative (e.g. Reinhard et al., 2017), where life is present, but its mark on the environment is unseen. Similarly the field of exoplanets has revealed stars and worlds, with evolutionary sequences that are likely very different to the Earth's around the Sun. By modeling these environments and their interactions, researchers have opened up the possibility of what was previously unthinkable for an Earth-like environment—that $O_2$ could be generated in a planetary environment without the action of life (e.g. Wordsworth & Pierrehumbert, 2014; Luger & Barnes, 2015; Tian et al., 2015; Domagal-Goldman et al., 2015; Gao et al., 2015; Narita et al., 2015; Harman et al., 2016) .

Our improved understanding of false negatives for $O_2$ may help us optimize target selection for the search for life on exoplanets, and it points to a rich area for future work. Being able to assess a potential target's false negative potential could be an integral part of the search for life, in addition to being crucial for interpreting what we do (and don't) detect. To advance this area of research, we will need to develop an improved understanding of planetary and stellar characteristics that may lead to false negatives, and the observational discriminants that might reveal them. As specific examples, $O_2$ may have only overwhelmed (or outlasted) its environmental sinks on Earth fairly recently, and so perhaps it would be prudent to prioritize older planets to allow time for a photosynthesis process to develop and leave its footprint. The interior and crustal composition of the planet may make it more likely for reducing gases to be present. Understanding how observed stellar composition affects terrestrial planet composition, how planetary orbit and star may affect tidal heating and surface heat flow, and searching for the presence of reducing gases or tectonic/volcanic activity via atmospheric gases or aerosols, may all be important for constraining possible sinks for $O_2$. Similarly, understanding the incoming stellar UV spectrum and activity levels and the overall impact on atmospheric chemistry and photolysis rates would also help identify processes that may suppress the detectability of a biosignature. In this example, young, volcanically active planets with strongly reducing atmospheres may not represent good targets for the search for oxygenic photosynthesis.

Our improved understanding of the star-planet processes that may lead to false positives for $O_2$, has only strengthened the robustness of $O_2$ as a biosignature. This knowledge can also inform our target selection, and prepare us to observe planetary characteristics that will help discriminate whether observed $O_2$ has a biological or abiotic origin. We now know of several abiotic planetary processes that will generate $O_2$, predominantly for planets orbiting M dwarfs (Wordsworth & Pierrehumbert, 2013; Tian et al., 2014; Domagal-Goldman et al., 2015; Luger & Barnes, 2015; Tian 2015; Harman et al., 2015). There are very likely more, and the search should continue for additional false positive mechanisms using modeling efforts or observations of exo-Venuses (Berta-Thompson et al., 2016; Gillon et al., 2017), that may provide observational tests for some of the proposed mechanisms. In particular, modeling is critically needed to anticipate the challenges we will face in interpreting the spectra obtained by future observatories.

Identification of potential false positives helps guide recommendations for the wavelength ranges and spectral resolutions needed to discriminate them, and allows us to draft a more sophisiticated and robust observing sequence for the search for life to exoplanets. We now know that the detection of any potential biosignature gas in an exoplanet's atmosphere must be considered in the context of the complex web of interrelated processes that shape the broader planetary environment. For example: Where is the planet relative to its star's habitable zone? What is the activity level of the star, and what types of photochemistry can it drive? Does the planet show other signs of habitability such as liquid surface water from water vapor or glint? What other gases are present? What fluxes are required to maintain the observed concentrations of the biosignatures gas? What false positive mechanisms may explain the observed possible biosignatures? Are there other potential biosignatures present?

If $O_2$ is detected, other characteristics of the planet can be observed to rule out false positive mechanisms, and to bolster the likelihood that the $O_2$ comes from a planetary biosphere. A larger age for the planet would again not only decrease the probability of a false negative, but decrease the likelihood of a false positive as well, as an initially high atmospheric $O_2$ abundance from ocean loss become sequestered or lost to space (Chassefière, 1996; Schaefer et al., 2016). Searching for a combination of gases or planetary characteristics, and not finding some of them, could also be powerful. Detecting $O_2$ and water or ocean glint, without $O_4$, can be more robust than $O_2$ alone. Detecting $O_2$, $CH_4$ and and water, or $O_2$ and water without strong $CO_2$ or CO, are both stronger detections. Finally, searching for secondary biosignatures of photosynthesis such as the red edge (Gates, 1985) and seasonal variability (Meadows, 2008), also strengthens the case for $O_2$ as a biosignature. Searching for the multiple planetary characteristics that strengthen the likelihood that O2 is a biosignature requires observations that span the broadest possible spectral range, and that are tailored for the strengths of the observing mode chosen, whether that is direct imaging, transmission, or high-resolution spectroscopy.

In closing, oxygenic photosynthesis evolved a metabolism that harvested light from the parent star to split water and fix carbon, and it ultimately rose to sculpt the history of our planet and produce the strongest biosignatures on Earth today. However, the absence of $O_2$ throughout much of Earth's history also encourages us to learn more about biosignatures that are not tied to oxygenic photosynthesis, to be better prepared for environments in which oxygenic photosynthesis has either not evolved, or not yet overwhelmed sinks for $O_2$.   Yet for any new biosignatures that are identified, the same process that has been developed for the study of $O_2$ should still be applied.   Biosignatures cannot exist distinct from the environment that harbors them, and understanding the impact of the environment on the detectability of biosignatures, and the potential generation of false positives, will be an important activity for future biosignature research.

## 7.0 Conclusions

The study of $O_2$ has catalyzed the interdisciplinary synthesis of research in early Earth science with modeling of star-planet interactions in exoplanet science has greatly expanded our understanding of biosignature science.   The early, simplistic view that $O_2$ alone constituted the most robust biosignature for detection of life on exoplanets has given way to a more sophisticated understanding of the impact of a planetary environment on the detectability and interpretation of $O_2$ in a planetary spectrum.  The delay between the advent of oxygenic photosynthesis on Earth, and the rise of atmospheric $O_2$ to modern levels, reveals a false negative process where life is present, but not detectable, due to suppression of the biosignature by the planetary environment.   Future research into false negative processes could reveal planetary processes and observational discriminants that could help identify planets on which biosignatures are likely suppressed.   This could potentially support exoplanet target selection for biosignature searches, and inform the interpretation of observed planetary spectra.   Similarly, the study of false positives has revealed stellar and planetary characteristics that may cause $O_2$ to build up abiotically in a planetary environment, and identified observational discriminants for those processes.   This allows observations of $O_2$ in a planetary spectrum to be more robustly interpreted as a biosignature by searching for and ruling out false positive mechanisms.  The processes to identify false negatives and positives for $O_2$ serve as an exemplar for a more generalized process for biosignature detection that should be applied to other novel biosignatures.

## Acknowledgments

This work would not have been possible without the collaborative community interactions fostered at  the Exoplanet Biosignatures Workshop organized and supported by the NASA Nexus for Exoplanet System Science. VSM, GNA, MNP, SD-G, APL, JLY and RD were supported to do this work by NASA Astrobiology Institute's Virtual Planetary Laboratory under Cooperative Agreement Number NNA13AA93A.